\documentclass{aa}
\usepackage{graphicx,comment}
\usepackage{txfonts}
\usepackage{footnote}

\usepackage{longtable}
\makesavenoteenv{tabular}
\makesavenoteenv{table}
\usepackage{tabularx}
\usepackage{diagbox}
\usepackage[colorlinks=true,
            linkcolor=red,
            urlcolor=blue,
            citecolor=blue]{hyperref}

\makeatletter
\renewcommand*\aa@pageof{, page \thepage{} of \pageref*{LastPage}}
\makeatother

\newcommand\clearrow{\global\let\rowmac\relax}
\clearrow
\begin{document}

   \title{The first look at Narrow-Line Seyfert 1 Galaxies with eROSITA
    }

\titlerunning{SDSS DR12 NLS1s in eRASS1}

   \authorrunning{Gr\"unwald, et al.}

   \author{G. Gr\"unwald\inst{1,2}\thanks{E-mail: grunwald@mpe.mpg.de,gruenwald.gudrun@gmail.com},
            Th. Boller\inst{1},
            S. Rakshit\inst{3},
            J. Buchner\inst{1},
            Th. Dauser\inst{4},
            M. Freyberg\inst{1},
            T. Liu\inst{1},
            M. Salvato\inst{1}, and
            A. Schichtel\inst{1,2}
          }

   \institute{Max-Planck-Institut f\"ur extraterrestrische Physik, Giessenbachstrasse 1, 85748 Garching, Germany
   \and
   Johann Wolfgang Goethe-Universit\"at Frankfurt, Max-von-Laue-Str. 1, 60438 Frankfurt am Main, Germany
   \and
   Aryabhatta Research Institute of Observational SciencES (ARIES), India
   \and
   Universit\"at Erlangen/N\"urnberg, Dr.-Remeis-Sternwarte, Sternwartstraße 7, 96049 Bamberg, Germany
            }

   \date{Received July 28, 2022; accepted November 9, 2022 }

  \abstract
  {
   {We present the first look at the spectral and timing analysis of Narrow-Line Seyfert 1 Galaxies (NLS1s) with the
   extended ROentgen Survey with an Imaging Telescope Array (eROSITA) on board the Spectrum-Roentgen-Gamma (SRG) mission. The sample of $\sim$1,200 NLS1s was obtained via cross-match of the first eROSITA All-Sky Survey (eRASS1) catalogue with the catalogue of spectroscopically selected NLS1s from
   the Sloan Digital Sky Survey (SDSS)
   DR12 by Rakshit et al. [ApJS, 229, 39 (2017)]. The X-ray spectral analysis is based on a simple power-law fit.
   The photon index distribution has a mean value of about 2.81$\pm$0.03, as expected from previous X-ray studies of NLS1s.
   Interestingly, it is positively skewed, and about 10 percent of the sources are located in the super-soft tail of photon indices larger than 4. These sources are of further interest as their source counts run into the X-ray background at values at around 1 keV.
   We argue that ionised outflows have been detected by eROSITA and may account for some of the extreme spectral steepness, which is supported by correlations found between the photon index and optical outflows parameters.
   We analysed the intrinsic X-ray variability of the eRASS1 to eRASS3 light curves of the sample but do not find significant variability neither during the individual survey scans nor between them.
   }

}


\maketitle
%

\section{Introduction}

The mass of the supermassive black hole at the centre of a galaxy is known to be strongly correlated with the stellar velocity dispersion, which is believed to be a consequence of the feeding and feedback working together to self-regulate the black hole growth and galaxy evolution
\citep{2013Kormendy}.
While the feeding mechanism provides a vast amount of gas supply to the supermassive black hole, helping the Active Galactic Nuclei (AGN)
to grow, the strong radiation from the AGN quenches the gas supply and halts the growth of the supermassive black hole, thereby regulating the black hole growth and galaxy evolution
(see \cite{2015King}).
The gas outflow, which is considered a tracer of feedback from AGN, has been reported at various wavelength bands e.g.\ optical, UV, and X-ray
(e.g.\ \cite{2003Reeves,2013Arav}),
allowing us to study feedback at different spatial scales. For example, kpc-scale outflow as traced by the asymmetry and velocity shift of the
[O III $\lambda$5007]
(e.g.\ \citet{2018Rakshit})
allows us to investigate the interaction of outflow with the interstellar medium  while the small-scale outflow as seen in the X-ray band could help to study the wind from the accretion disk
(e.g.\ \cite{2003ARA&A..41..117Crenshaw,2009ApJ...701..493Reeves,2015Natur.519..436Tombesi}).
The previous studies suggest strong correlations between
[O III $\lambda$5007]
velocity shift and dispersion with the accretion rate of AGN, suggesting the gas outflow is driven by the accretion rate
\citep{2018Rakshit}.
Especially for objects with high accretion rates, AGN feedback processes are thought to be of importance.
Outflowing winds launched from the accretion
disc by radiation pressure or magnetic
fields are considered an important AGN feedback process.
For radiation-pressure-dominated winds, outflows can reach velocities up to about 0.3\,c and can drive substantial amounts of material into the interstellar medium. These winds have been discovered so far mainly based on XMM-Newton observations
(e.g.\ \citealt{2003Poundsa, 2003Poundsb, 2003Poundsc, 2003King, 2003Reeves}).
Outflowing winds
with such high velocities have been named
Ultra-fast outflows (UFOs)
by \cite{2010Tombesi} in a systematic study of bright XMM-Newton AGN.
Reviews on AGN outflows in general have been presented by e.g.\ \citep{2012Fabian, 2021Parker}.

Narrow-line Seyfert 1 (NLS1) galaxies are a subclass of AGN with high Eddington rate and are defined by the relative narrowness of their broad emission lines, i.e.\ FWHM of the broad H$\beta$ line being less than $\sim 2000$\,km/s,  a ratio of the [OIII] to H$\beta$ line flux of less than 3, and strong Fe II multiplet emission \citep{OP1985,Goodrich1989}.
NLS1s are thought to be AGN with low black hole masses in their early phase of accretion (e.g.\ \cite{1996Boller}). Due to their high Eddington rates, and strong outflows, they are ideally suited for the study of accretion physics in the strong GR regime in the innermost regions of AGN, and relativistic effects are expected to be prominent features of eROSITA {\citep{2012Merloni,2021Predehl}} observations.
Therefore eROSITA observations are
expected to advance our understanding of matter under strong gravity, and determining the NLS1 content in the eROSITA data will build upon our present understanding of the Seyfert phenomenon.
%

eROSITA, developed by MPE and launched in 2019, now provides the first all-sky survey in the complete 0.2-8\,keV band with better energy resolution compared to ROSAT and comparable sensitivity in the soft X-rays compared to XMM-Newton.
In this paper the first population study on a large homogeneous sample of $\sim$1,200 spectroscopically confirmed NLS1s was performed using eROSITA data, marking a 25-fold increase in sensitivity and much improved spectral resolution in the soft X-ray compared to
the ROSAT
survey
data used in previous studies (e.g.\ \cite{2007Anderson}, \cite{2018Rakshit}).
The unique soft X-ray energy response of eROSITA
allows for the first time studies of the soft X-ray spectral energy distribution on the largest sample of spectroscopically confirmed NLS1s and to study outflow signatures from the gravitational radius scale to kpc-scale.

The structure of the paper is as follows.
We describe our sample selection in section \ref{sec:sample}, the X-ray spectral analysis is given in section \ref{sec:x-ray}.
The optical line asymmetry and outflow indicator parameter determination is outlined in section \ref{sec:optical_spectra}.
In section \ref{sec:results} we present the results or this paper, the variability results are given in section \ref{sec:timing}, and section \ref{sec:discussion} contains the discussion part. We summarise our results in section \ref{sec:summary}.

\section{DR12 NLS1s cross-match with eRASS1}\label{sec:sample}

eROSITA's main operational phase is the 8-fold all-sky survey, labelled eRASS1--8.
A simple positional cross-match of eRASS1 X-ray counterparts using astrometrically corrected positions
to the optical coordinates of the
Sloan Digital Sky Survey (SDSS)
DR12 catalog of  \citet{RS017} within 12 arcsec
yielded 2,238 matched sources.

The source products have been obtained by the eROSITA Science Analysis Software System (eSASS) configuration which is designed for eRASS1 data analysis and source products and corresponding data release \citep{Merloni2023}.

Sources for which less than 10
net photon counts had been detected were rejected from the spectral fitting analysis due to expected insufficient signal-to-noise, leaving
1,374 sources, for
1,197 spectral files were available produced by the eSASS pipeline.
%
We note that the combined positional uncertainty of the eRASS1 and the DR12 surveys
peaks around 6 arcsec.
%
Choosing the maximum error to be 12 arcsec for unambiguity of the correlation only led to minimal losses. Also, note that the projected proximity of two detections does not automatically mean the sources are identical. More sophisticated cross-matching algorithms taking into account magnitudes, colours, and other additional information such as NWAY (e.g.\ \cite{Salvato2018}) are available. In this case, the assignment of the counterparts by NWAY compared to the simple positional cross-match differed only marginally.

\begin{figure}[htp]
\includegraphics[width=\columnwidth]
{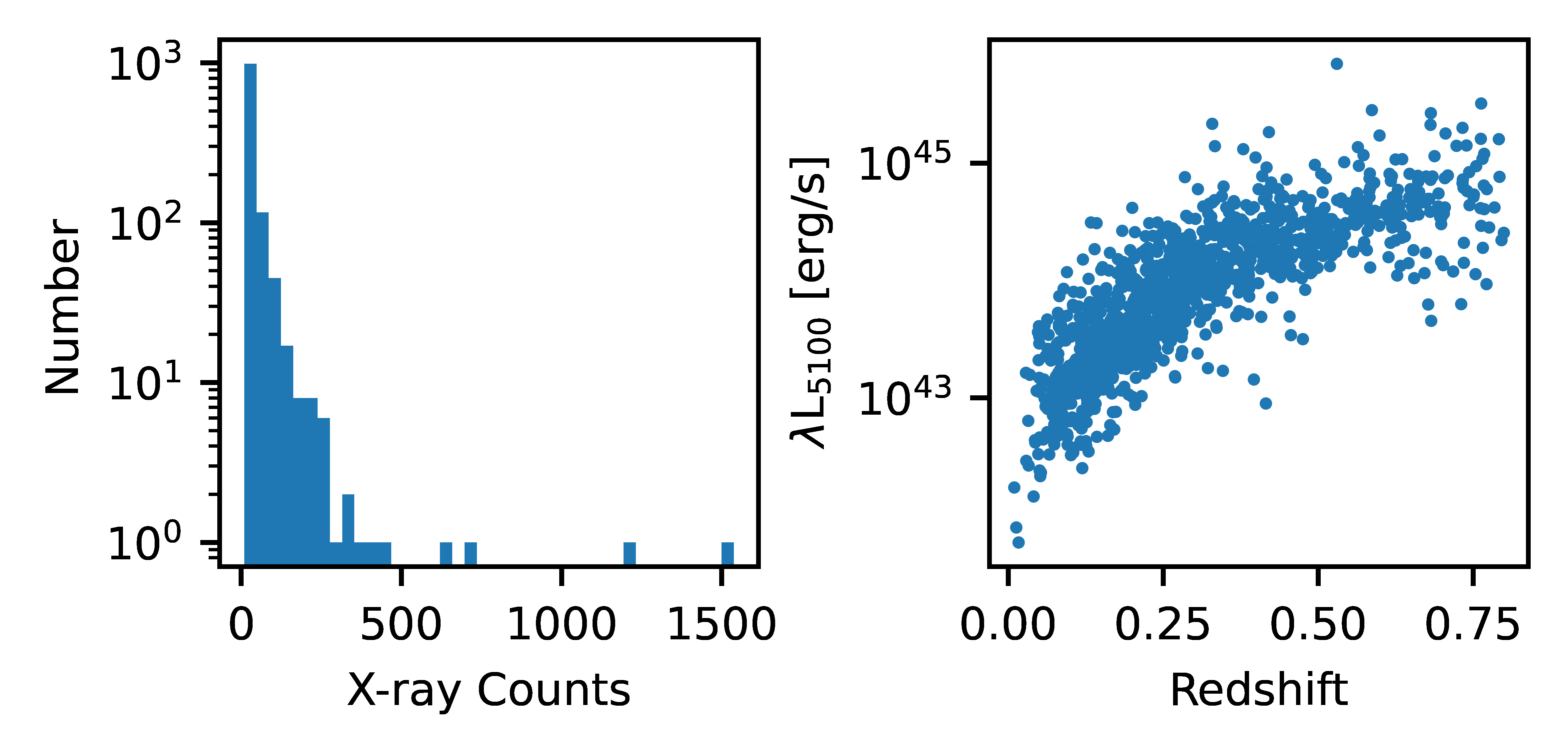}
    \caption{$\lambda\rm L_{5100}$, redshift and X-ray counts distributions of the eRASS1 NLS1 sample.
    }
    \label{histo_z_counts}
\end{figure}

The $\lambda\rm L_{5100}$ versus redshift distribution of DR12 - eRASS1 NLS1 and the distribution of X-ray counts are shown in Fig.~\ref{histo_z_counts}.

\section{X-Ray spectral analysis}\label{sec:x-ray}

PyXspec was used for automatised maximum-likelihood-based fitting using the Cash statistics, the modified Levenberg-Marquardt algorithm, and a Principal Component Analysis (PCA)-based automated background modelling tool implemented in the Bayesian X-ray Analysis (BXA) software package \citep{Buchner2014} as described in \citet{Simmonds2018}
to obtain the spectral properties of the eRASS1 NLS1 cross-matched sample
as described in
\cite{Gruenwald2022}.

A combined black body and powerlaw model to account for the soft X-ray excess emission from the disk and the coronal emission, respectively, was tested.
Due to the relatively low photon count statistic resulting from the short exposure times, the parameter spaces appeared to be very inhomogeneous and featured multiple local minima in the physically meaningful parameter ranges, some of which were not physically plausible, so that fits performed with the algorithm at hand would have to be deemed unreliable \citep{Gruenwald2022}.

Instead, a simple power-law model (\texttt{zpowerlw} in \texttt{XSPEC}) with foreground absorption (\texttt{TBabs} with the \citet{Verner1996} photoionisation cross sections and the
\citet{Wilms2000} abundances) was adopted. The corresponding equivalent hydrogen column
density (in
units of $10^{22}$\,atoms/cm$^2$) $N_\mathrm{H}$ was initially kept as a free parameter; only if
$N_\mathrm{H}$ 
could not be fitted it was fixed to the Galactic value \footnote{The galactic column was obtained from \url{https://www.swift.ac.uk/analysis/nhtot/index.php}}
in 
the direction of the source\textcolor{red}{,} $N_\mathrm{H, gal}$.

\begin{figure}[htp]
\includegraphics[width=\columnwidth]{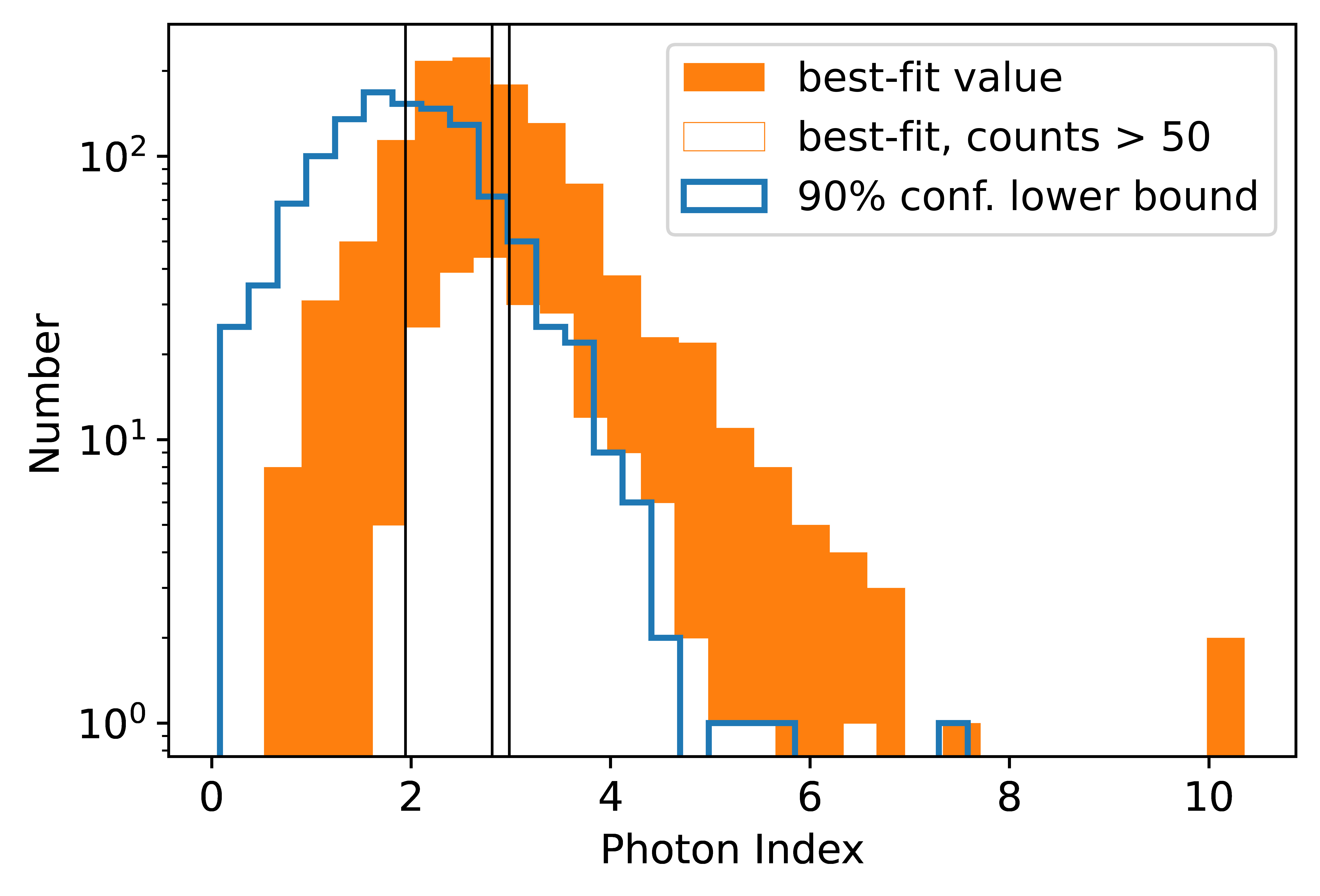}
    \caption{
    Photon index distribution of the NLS1 DR12 eRASS1 sample as obtained from a simple power-law fit when freezing $N_\mathrm{H}$ only if necessary to obtain a valid fit (filled). The mean photon index is $\rm 2.81\pm0.03$. About 10 percent of the sources are located in the super-soft tail with photon indices between 4 and $\sim$10. The white within the filled histogram is the best-fit value for sources with X-ray counts greater than 50 (mean: $2.99\pm0.05$), and the line is the distribution of the 90\% confidence lower bounds with a mean of 1.94$\pm$0.02.
    }
        \label{Gamma-histo}
\end{figure}

The photon index distribution obtained from the power-law fit is shown in Fig.~\ref{Gamma-histo}. The mean photon index
and the standard error of the mean are $\rm 2.81\pm0.03$.
%
%

This is steeper compared to other AGN samples obtained with eROSITA, e.g.\ \citet{Liu2021} as expected for NLS1s. 
We note that about 10\% of the distribution has photon indices above 4, reaching values up to about 10. This super-soft tail has not been detected to such an extent in previous X-ray NLS1 studies.

\begin{figure}[htp]
\includegraphics[width=\columnwidth]{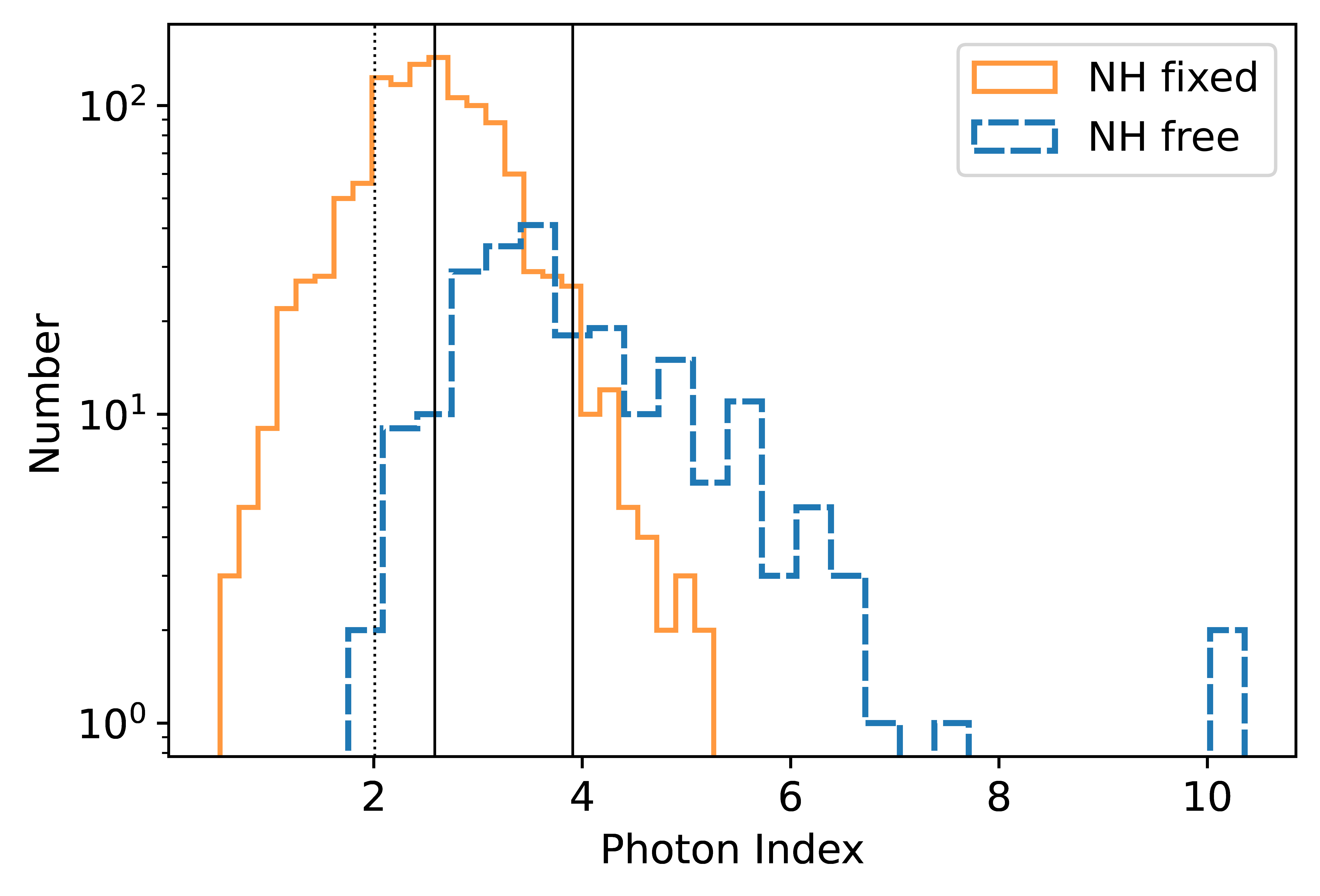}
    \caption{
   Solid line: Distribution of $\Gamma$ obtained with $N_\mathrm{H}$ fixed to the Galactic value for the entire sample. Dashed line: Histogram of $\Gamma$ for the subsample, for which the fit with $N_\mathrm{H}$ as a free parameter succeeded. The solid vertical lines mark the means at 2.6 ($N_\mathrm{H}$ fixed) and 3.9 ($N_\mathrm{H}$ free). For the latter subsample, the distribution is shifted to slightly higher photon indices. Some increase in steepness is realistic due to the degeneracy as $N_\mathrm{H, gal}$ gives a lower bound to the true value of $N_\mathrm{H}$.
   The dotted vertical line is the typical slope of X-ray emitting AGN in the eFEDS field as shown by \cite{Liu2021}.
    }
    \label{Gamma-histo_nhfixed}
\end{figure}

To account for the known degeneracy between photon index and $N_\mathrm{H}$ for low count spectra, we fixed $N_\mathrm{H}$ to the Galactic value for all sources. The resulting distribution has a slightly lower mean photon index with
$\rm 2.59\pm 0.02$, but still reaches values above 5 (Fig.~\ref{Gamma-histo_nhfixed}).
The mean values, errors, as well as the standard deviations of $\Gamma$ for the different treatments of $N_H$ as shown in Figures~\ref{Gamma-histo} and ~\ref{Gamma-histo_nhfixed} are listed in Table~\ref{diststatistic}.

\begin{table}
\centering
\begin{tabular}{l|lll}
{\diagbox{$N_H$~}{statistic of $\Gamma$}}                & mean & sem  & std   \\
\hline
fixed                                         & 2.59 & 0.02 & 0.71  \\
free                                          & 3.91 & 0.08 & 1.21  \\
mixed, best-fit value                         & 2.81 & 0.03 & 1.00  \\
mixed, best-fit value, counts  50             & 2.99 & 0.05 & 0.72  \\
mixed, 90\% confidence lower bound & 1.94 & 0.02 & 0.84
\end{tabular}
\caption{Mean value, standard error of the mean, and standard deviation of $\Gamma$ for the discussed treatments of $N_H$.}
\label{diststatistic}
\end{table}

\begin{figure}[h!]
\includegraphics[width=\linewidth]{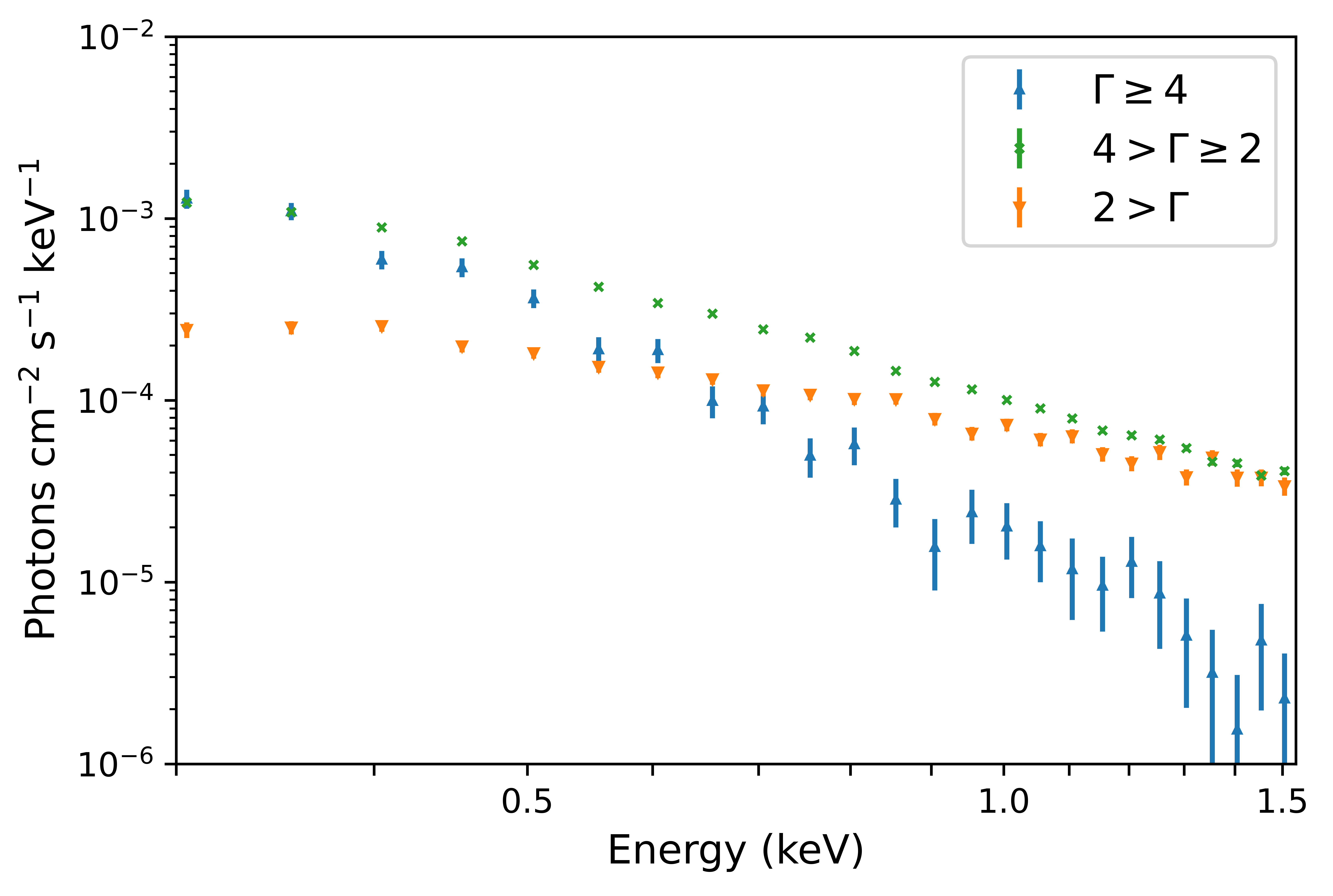}
    \caption{Averaged unfolded spectra of sources showing a high ($\Gamma \geq 4$), moderate ($4 > \Gamma \geq 2$), and low ($2 > \Gamma$) photon index, respectively, when fitted with $N_H$ fixed. For the gamma distribution of this fit, see Fig.~\ref{Gamma-histo_nhfixed}. For improved visualisation, the data was binned into bins with a width of 0.05 keV.
 }
        \label{fig:averaged_unfolded_spectra}
\end{figure}

In Fig.~\ref{fig:averaged_unfolded_spectra} we show the average unfolded spectrum for objects with $\Gamma \geq 4$, $4 > \Gamma \geq 2$, and $2 > \Gamma$, respectively, as determined by fitting the powerlaw component only. For the corresponding distribution of the photon index, see Fig.~\ref{Gamma-histo_nhfixed}. The observed spectra were unfolded using XSPECS \texttt{ufspec} command and the respective best-fitting parameters of the powerlaw model, and the stacking within each steepness group was done by taking the arithmetic mean of the unfolded data following the approach by
\cite{LiuZhu2016}.


%
%

%

\section{Partial covering by an outflowing ionised absorber}

\citet{Gruenwald2022} has shown that ionised outflowing winds can explain photon indices with values above 5.
As opposed to a cold (neutral) absorber, a warm absorbing material where lighter elements are ionised will be transparent to soft X-rays while still being opaque to higher energies. At the same time, it will imprint a prominent absorption feature composed of lines from moderately ionised Fe, Ne, and Mg. This causes a loss of photons beyond the soft band, manifesting as a steeper spectrum. As the photon count statistic of the eRASS1 data is not sufficient to directly test this model on the data, we simulate the scenario using XSPEC's \texttt{fakeit} command on their \texttt{zxipcf} model.

\begin{figure}[h!]
\includegraphics[width=\linewidth]{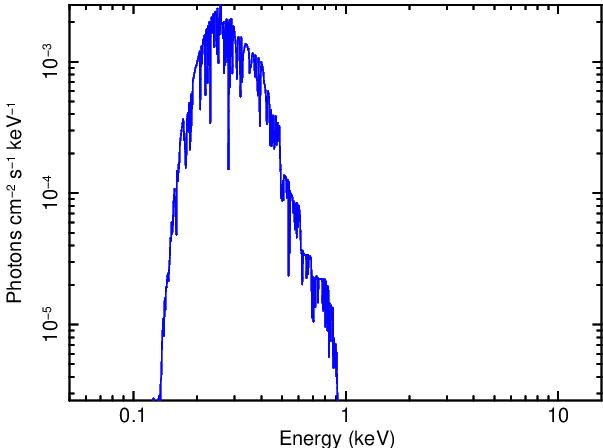}
\includegraphics[width=\linewidth]{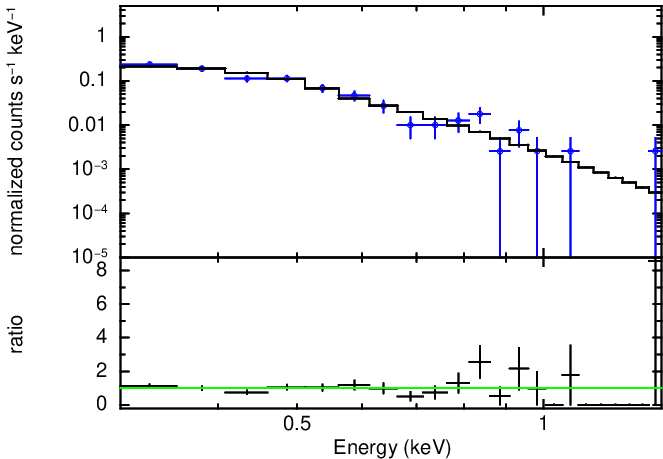}
    \caption{Top panel:
    Partially covering outflowing ionised absorber model (\texttt{zxipcf}) set up for the simulation with $NH=6 \times 10^{22}$, $\log(\xi)=1$, $f_{cov}=0.99$, $z_{outflow}=-0.2$, and a powerlaw source with  $\Gamma=4.5$, $z=0.3$, and $NH=0.04\times 10^{22}$.
    Bottom panel:
    Fitting the simulated data with a powerlaw gives $\Gamma=7.4$ when setting $NH=0.1\times 10^{22}$.
    }
        \label{fig:data}
\end{figure}

Figure \ref{fig:data} shows the effect of an ionised absorber on a power-law of $\Gamma=4.5$,
at a redshift of $z=0.3$ (representative for the sample)
with a galactic absorption of $NH=0.04\times 10^{22}$.
The absorber is given a column density of $NH=6 \times 10^{22}$, an ionisation parameter of $\log(\xi)=1$\footnote{The ionisation parameter $\xi$ is defined as $\xi = \frac{L}{nr^2}$, where $L$ is the ionisation luminosity, $n$ is the gas density, and $r$ is the distance between the source and the absorber.}, and a very high covering fraction of $f_{cov}=0.99$.
Note that a change in covering fraction could lead to an observed variability of the X-ray flux.

The absorber is blue-shifted towards the observer with $z=-0.2$, corresponding to an outflow velocity of $\sim$0.4~c.
Fitting the data simulated for this scenario from 0.3 to 1.5~keV with a powerlaw while setting $NH=0.1\times 10^{22}$, as the best fits of the steep objects of the sample suggest, a photon index of $\Gamma=7.4$ is reached\footnote{When keeping the initial value for $NH$, the photon index is increased to $\Gamma=6.9$, compared to the initial value of $\Gamma=4.5$.}, 
which further increases when assuming even higher outflow velocities.



\begin{figure}
    \centering
\includegraphics[width=9.0cm]{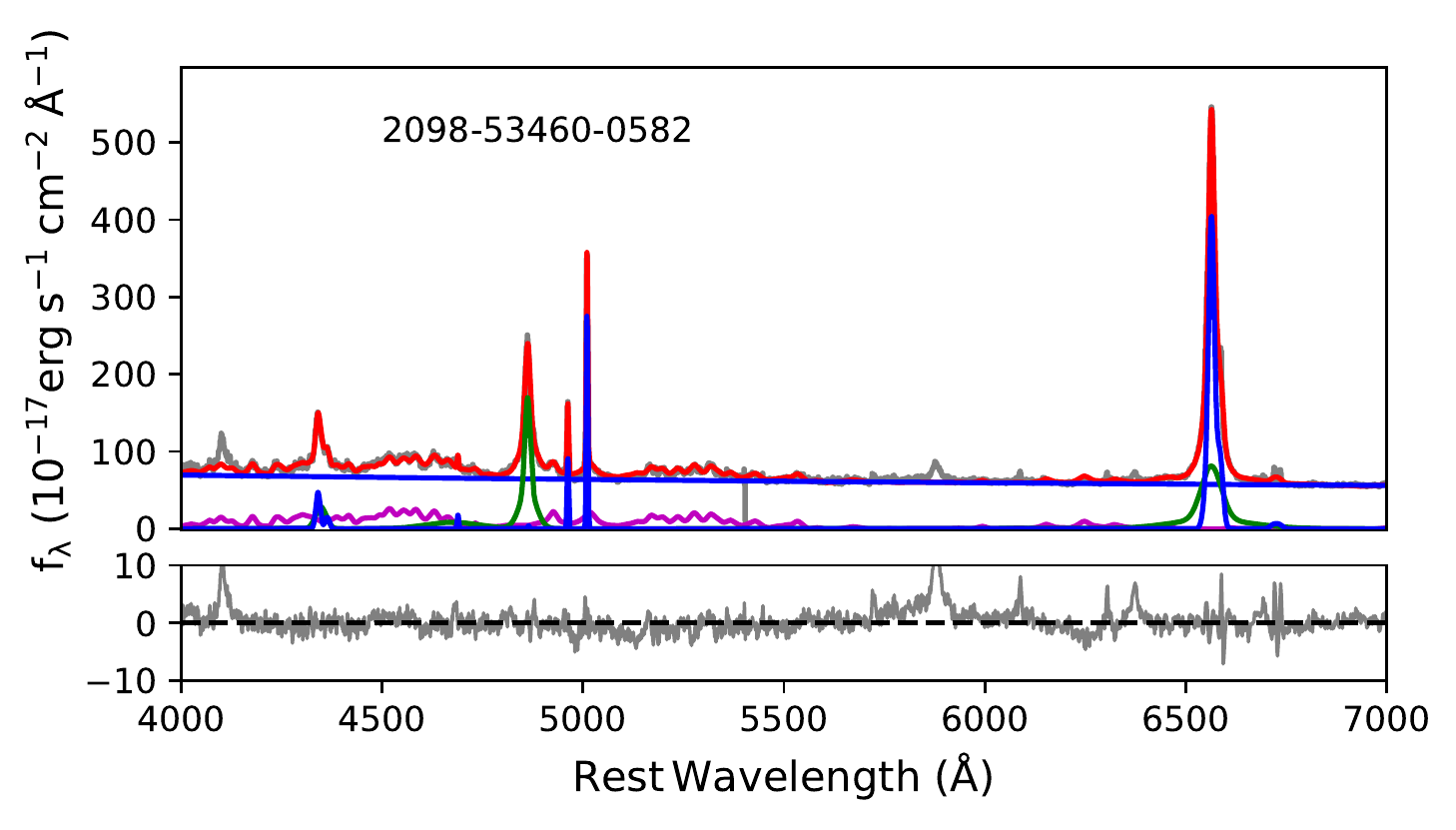}
    \caption{An example of optical spectral fitting. The data (black), best-fit model (red), and individual decomposed components i.e., the power-law (blue), Fe II (magenta), broad line (green), and narrow line (blue) are shown for a NLS1 with SDSS ID of 2098-53460-0582. The lower panel shows the residual in the unit of error spectrum.}
    \label{fig:sdss_spectra}
\end{figure}

\section{Optical spectral analysis}\label{sec:optical_spectra}

In order to estimate the line asymmetry and the outflow indicators such as velocity shift and dispersion of [OIII] line, we fitted all the spectra carefully to estimate various properties that are not available in the catalog given in
\citet{2017Rakshit}. For this purpose we used
the
publicly available code \textsc{PYQSOFIT} developed by \cite{2018ascl.soft09008G}. A detailed description of the code and the spectral fitting method used in the paper can be found in \cite{2020ApJS..249...17R}. We corrected each spectrum for Galactic extinction using the \cite{1998ApJ...500..525S} map and extinction law from \cite{1999PASP..111...63F} for
the Milky Way
with
$R_v = 3.1$.

We de-redshifted each spectrum using the SDSS redshift and then performed detailed modelling of each spectrum. In short, we first decomposed stellar light using the PCA method from each spectrum using 5 PCA components for galaxies and 20 PCA components for quasars that can reproduce about 98\% of the galaxy sample and 96\% of the quasar sample, respectively \citep{2004AJ....128.2603Y}. We then subtracted the stellar contribution and modelled the AGN continuum, masking strong emission
lines, 
as a combination of power-law for AGN contribution and Fe II emission using optical Fe II template from \cite{1992Boroson} and UV Fe II template from UV template built by
\cite{2019ApJS..241...34S}.

\begin{figure*}
\includegraphics[width=18.0cm]{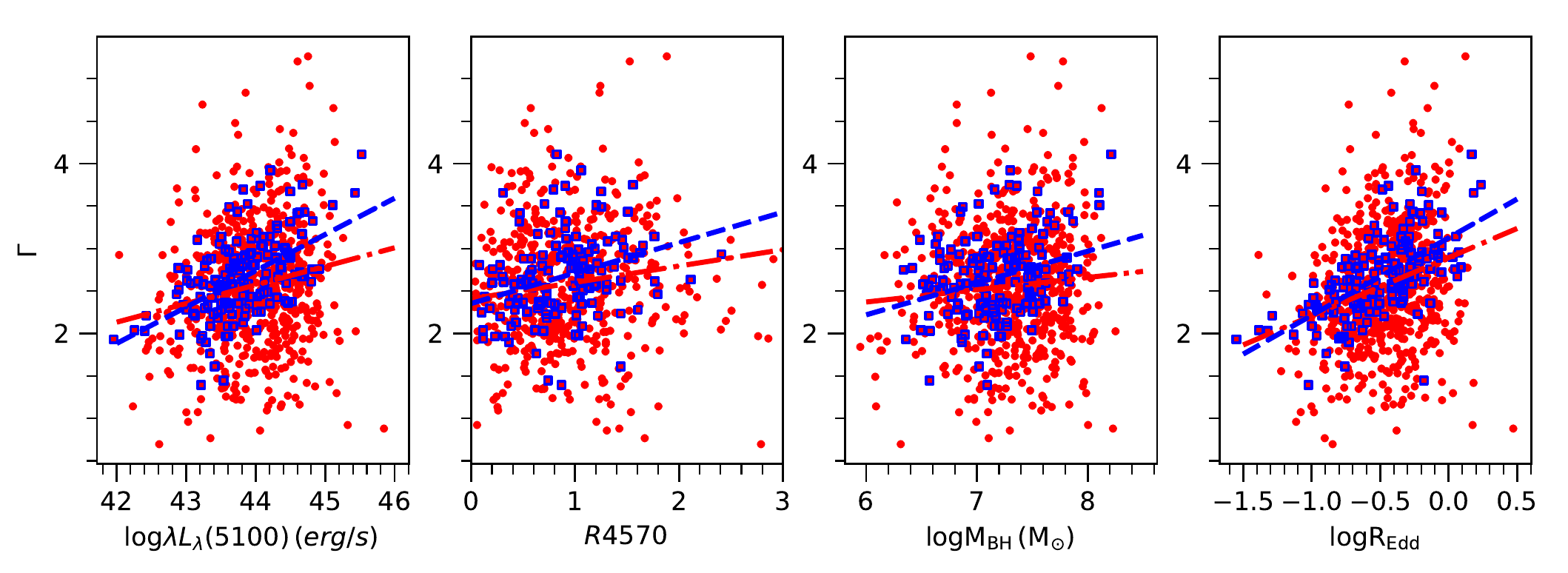}
\caption{Correlation of the photon index obtained for fixed $N_\mathrm{H}$ fit to SDSS DR12 sample properties (luminosity, Fe II strength, black hole mass, and Eddington ratio) for sources with photon counts larger than 10 (red-dot) and 50 (blue-square). The best-fit lines for both samples are shown.}
\label{Gamma_params_corr}
\end{figure*}

To get the pure emission line spectrum and model the emission lines, we subtracted the best-fit continuum model from the spectrum and then performed multi-component modelling in individual line complexes. The broad emission lines (H$\beta$ and H$\alpha$) were modelled as a combination of broad components using two Gaussian's each having FWHM $>900$ km/s and a narrow component using a single Gaussian of FWHM$<900$ km/s. We modelled the narrow lines using a single Gaussian of FWHM $<900$ km/s except the [O III $\lambda$5007] which we modelled using two Gaussian's, one for the wing and one for the core components. During the fitting following restriction was applied F(5007)/F(4959) = 3, F(6585)/F(6549) = 3, and velocity shift and width of narrow lines in a given complex are tied to each other. An example of spectral fitting along with the decomposed components is shown in Fig.~\ref{fig:sdss_spectra}.

We estimated the uncertainty in each derived quantity via 
Markov chain Monte Carlo (MCMC) adding Gaussian noise based on the flux uncertainty to each spectrum and repeating this for 50 iterations. From the distribution of each quantity, we estimated 1$\sigma$
uncertainties 
on both sides (covering 16\% and 84\% of the sample) around the median.

Along with emission line width and flux of several emission lines, we also estimated the asymmetry index (AI) to characterise the shape of the emission lines using the method described in
%
\citet{1981Heckman,1996Marziani,2020Wolf}.
We used 75\% and 25\% flux values for the calculation of AI using the following equation.

\begin{equation}
    AI = \frac{\lambda_{\mathrm{red}} + \lambda_{\mathrm{blue}} - 2\lambda_{\mathrm{peak}} }{\lambda_{\mathrm{red}} - \lambda_{\mathrm{blue}} }
\end{equation}
where $\lambda_{\mathrm{peak}}$ is the wavelength corresponding to the peak flux, $\lambda_{\mathrm{red}}$ and $\lambda_{\mathrm{blue}}$ are the wavelengths corresponding to the red and blue wings where the flux reaches 25\% of the peak flux. For broad Balmer lines, we used only the broad component and for [O III $\lambda$5007], we used the total profile to calculate the AI.

The line profile of [O III $\lambda$5007] has been used in literature to study outflow on the kpc scale, e.g.\
%
\citet{2018Woo,2018Rakshit}.
The [O III $\lambda$5007] kinematics are mostly driven by non-gravitational components due to outflow, although the virial motion due to the gravitational potential adds to the line broadening.

The velocity dispersion ($\sigma_{\mathrm{line}}$) and shift of the [O III  $\lambda$5007] line are useful indicators of the outflow and can be estimated as
\begin{equation}
    \sigma^2_{\mathrm{line}} = \frac{\int \lambda^2 f_{\lambda} d \lambda }{\int f_{\lambda} d \lambda} - \lambda^2_{\mathrm{avg}}
\end{equation}
where \begin{equation}
    \lambda_{\mathrm{avg}} = \frac{\int \lambda f_{\lambda} d \lambda }{\int f_{\lambda} d \lambda}
\end{equation}

We calculated the velocity shift of [O III $\lambda$5007] (V[O III]) with regard to systemic velocity, which could be measured with respect to the shift of the H$\beta$ narrow component
\cite{2018Rakshit}.
In order to reliably estimate the above quantities, a strong
[O III $\lambda$5007]
line is required. We, therefore, restricted further analysis  with S/N at
[O III $\lambda$5007]
larger than 5 and H$\beta$ broad component detected at 5$\sigma$. This restricts us to a sample of 723 objects for further analysis which is presented in the next section.

\begin{figure}[htp]
\includegraphics[width=9.2cm]{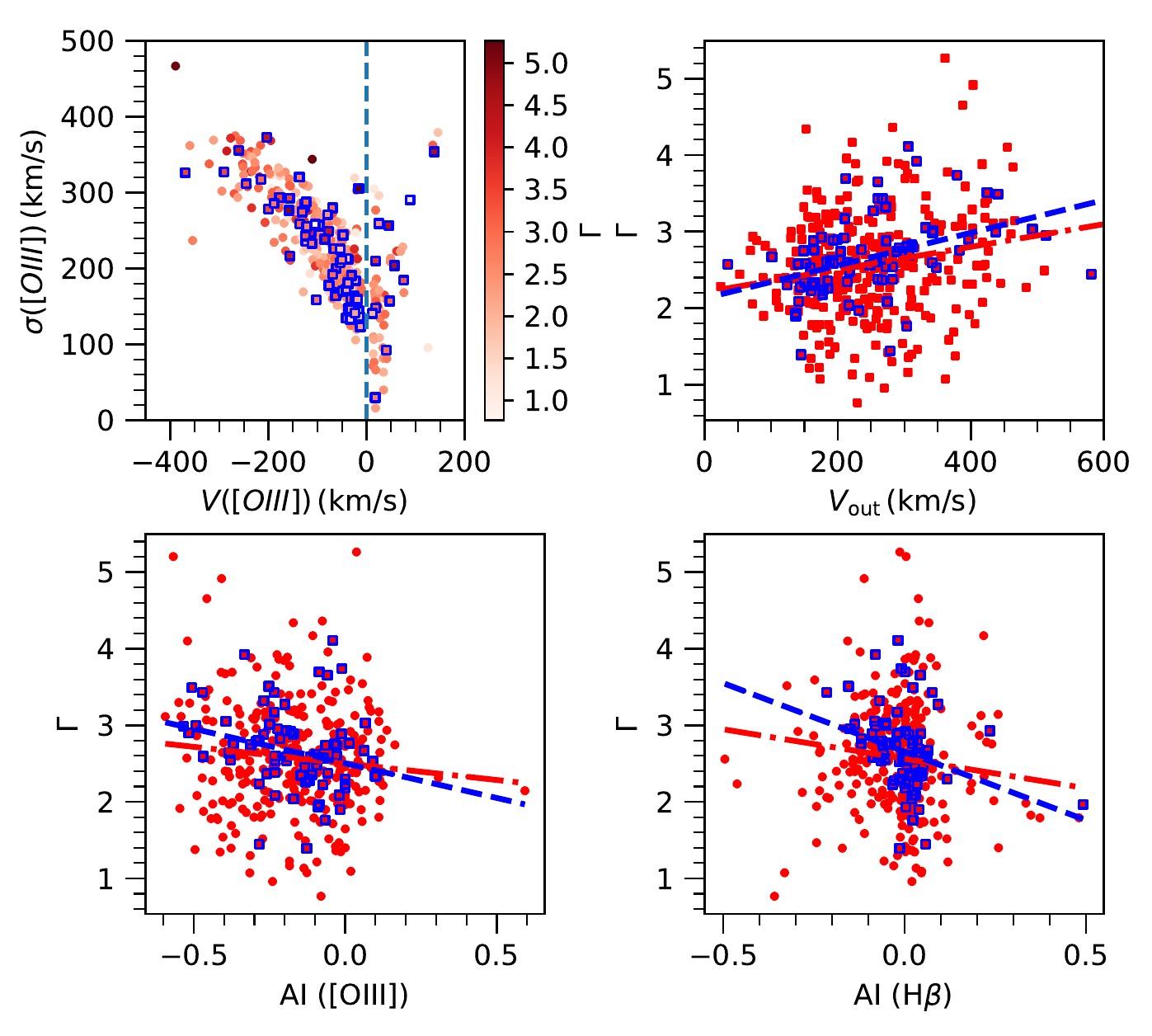}
\caption{VVD diagram and correlation between photon index and outflow velocity ($v_{out}= \sqrt(v[oiii]^2 + \sigma[oiii]^2)$) for fixed $N_\mathrm{H}$. The objects with zero velocity and zero AI of
[O III $\lambda$5007]
have been removed. Bottom: Correlation of the photon index with outflow parameters for fixed $N_\mathrm{H}$ for sources with photon counts larger than 10 (red-dot) and 50 (blue-dot). The best-fit lines for both samples are shown.}
\label{gamma_AI_corr}
\end{figure}

\section{Results}\label{sec:results}
\subsection{Dependency of photon index with AGN parameters}
To study the dependency of photon index with the physical parameters of AGN, we have calculated black hole mass based on the virial relation
\begin{equation}
    M_{\mathrm{BH}} = f R_{\mathrm{BLR}} \Delta V^2/G
\end{equation}
where $R_{\mathrm{BLR}}$ is the size of the broad line region (BLR) estimated using the BLR size-luminosity relation from \cite{2013Bentz}.
$\Delta V^2$ is the line FWHM and  f is the virial scale factor which is taken to be 1.12 from \cite{2015Woo}.
We also estimated Eddington ratio ($R_{\mathrm{Edd}}$) which is the ratio of bolometric ($L_{\mathrm{bol}}$) to Eddington ($L_{\mathrm{Edd}}$) luminosity, where the bolometric luminosity is estimated as $L_{\mathrm{bol}} = 9\times L_{5100}$ and $L_{\mathrm{Edd}}$ is estimated as $L_{\mathrm{Edd}} = 1.26\times10^{38}M_{\mathrm{BH}}$.

\cite{1992Boroson} studied 87 PG quasars and introduced the AGN Eigenvector-1 (E1) space which can explain diverse properties of broad-lined AGN. The EI space has been extended in different wavelength bands and the current E1 space is dominated by three parameters photon index in soft-X-ray, R4570, and FWHM of the line.

In Fig.~\ref{Gamma_params_corr} we show the correlations between the photon index against $L_{5100}$, Fe II strength (R4570) which is the flux ratio of Fe II (4435-4685 \AA) to H$\beta$ broad component, $M_\mathrm{BH}$ and $R_{\mathrm{Edd}}$. The Spearman's rank correlations parameters and p-value of no-correlation are listed in Table~\ref{tab:corr}. We also performed linear fits to the correlation between $\Gamma$ and various parameters using LINMIX code\footnote{\url{https://github.com/jmeyers314/linmix}} \citep{2007ApJ...665.1489K} with intrinsic random scatter ($\epsilon$) about the regression (here $\epsilon$ is assumed to be normally distributed with zero mean and variance ${\sigma^2}_{\mathrm{int}}$) and noted the results in Tab.~\ref{tab:linmix}.

The strongest correlation between these parameters is the positive correlation between photon index $\Gamma$ vs.\ $R_{\mathrm{Edd}}$, which agrees well with the findings in the literature. For example, \cite{Ojha2020} found a strong negative correlation between two parameters considering the soft X-ray photon index of an NLS1 sample. This correlation is also present in the total sample of NLS1 (221 objects) and Broad-Line Seyfert 1 Galaxies (BLS1) (154 objects) sample based on ROSAT and XMM-Newton spectral fitting. \cite{2016Wang} studied the correlation between the hard X-ray (2-10\,keV) photon index and various physical parameters and found a strong positive correlation between photon index and $R_{\mathrm{Edd}}$. Therefore, this correlation is not only valid for soft X-rays but also for the hard X-ray photon index.

We note that $R_{\mathrm{Edd}}$ is $\propto$ $L_{5100}$ and inversely to Eddington luminosity or the black hole mass, which is again proportional to $L^{1/2}_{5100}$ and the square of the $\Delta v$. We found a strong positive correlation between $\Gamma$ and $L_{5100}$ and no correlation between $\Gamma$ and $M_{BH}$. The latter is due to the combined effect of $\Gamma-L_{5100}$ correlation and the anti-correlation between $\Gamma$ and $\Delta v$, which is previously shown by various authors \citep[e.g.,][]{2016Wang,Ojha2020}), especially prominent for a sample with a large range of $\Delta v$. The Fe II strength and Eddington ratio are known to be correlated suggesting a positive correlation between $\Gamma$ and R4570 is expected, which is also found in our eRASS1 sample.

The correlation coefficient found in this large sample eRASS1 study is lower than the one found in previous studies based on a much smaller sample. Note that our original sample considers all the objects with photon counts larger than 10\footnote{Throughout the paper, we used sources with net photon counts larger than 10 unless mentioned otherwise.}. Considering the sources with large photon counts could improve the correlation statistics. To test this, we performed the correlation at various photon counts level e.g., $>30$ and $>50$. We found a much stronger correlation when objects with higher photon counts are considered (see values in parenthesis in Tab.~\ref{tab:corr} and the blue points in
Fig.~\ref{Gamma_params_corr} for sources with photon counts larger than 50), however, this reduced the sample size significantly as shown in
Fig.~\ref{histo_z_counts}. Therefore, the correlations become weaker when sources with low photon counts are considered.



\begin{table}
\resizebox{1.0\linewidth}{!}{
    \centering
\begin{tabular}{l l l l l l l} \hline  \hline
Col. & y    &  x         & $r_s$ & p-value & $r_s$ & p-value \\
     &      &            & \multicolumn{2}{ c }{$N_\mathrm{H}$ fixed} & \multicolumn{2}{ c }{$N_\mathrm{H}$ mixed}\\
(1)  & (2)  & (3)        & (4)   & (5)     & (6) & (7) \\ \hline
1   & $\Gamma$ & $\log L_{5100}$   & 0.20(0.53) & 5e-8(1e-12)  & 0.09(0.38)  & 8e-3(1e-6) \\
2   & $\Gamma$ & $R_{4570}$   & 0.10(0.32)   & 0.01(1e-4)   & 0.08(0.36)  & 0.02(1e-5) \\
3   & $\Gamma$ & $\log M_{BH}$     & 0.08(0.26) & 0.02(1e-3)   & -0.01(0.13) & 0.65(0.09) \\
4   & $\Gamma$ & $\log R_{EDD}$    & 0.28(0.65) & 2e-15(1e-19) & 0.22(0.50)  & 2e-9(5e-11) \\
5   & $\Gamma$ & v(OIII)      & -0.16(-0.29)  & 2e-3(9e-3)  & -0.14(-0.26) & 8e-3(0.02)    \\
6   & $\Gamma$ & $\sigma$(OIII)& 0.18(0.46)  & 4e-4(1e-5)   & 0.14(0.40)  & 8e-3(2e-4)    \\
7   & $\Gamma$ & AI ($OIII$)   & -0.10(-0.33) & 0.06(3e-3)  & -0.07(-0.21) & 0.16(0.06)    \\
8   & $\Gamma$ & AI (H$\beta$) & -0.11(-0.34) & 0.02(2e-3)  & -0.08(-0.30) & 0.11(7e-3) \\
9   & $\Gamma$ & $v_{out}$     & 0.20 (0.48) & 2e-4(6e-6)   & 0.16(0.42)  & 2e-3(1e-4) \\
10 & $\Gamma$ & $\log \dot{M}_{\mathrm{out}}$ & 0.22(0.46)& 1e-5(2e-5) & 0.14(0.34) & 6e-3(2e-3) \\
11 & $\Gamma$ & $\log \dot{E}_{\mathrm{out}}$ & 0.23(0.53)& 5e-6(3e-7) & 0.16(0.41) & 1e-3(1e-4) \\
12 & $\Gamma$ & $\log \dot{P}_{\mathrm{out}}$ & 0.23(0.51)& 7e-6(1e-6) & 0.15(0.40) & 3e-3(3e-4)\\
13 & $\log R_{EDD}$ & $\log \dot{M}_{\mathrm{out}}$ & 0.62 (0.68) & 6e-42 (5e-12) & 0.64 (0.68) & 2e-42 (1e-11)\\
14 & $\log R_{EDD}$ & $\log \dot{E}_{\mathrm{out}}$ & 0.58 (0.67) & 7e-36 (2e-11) & 0.59 (0.67) & 4e-36 (3e-11) \\
15 & $\log R_{EDD}$ & $\log \dot{P}_{\mathrm{out}}$ & 0.61 (0.69) & 4e-39 (3e-12) & 0.62 (0.68) & 2e-39 (6e-12) \\
 \hline
 \hline
    \end{tabular} }
    \caption{Correlation results of photon index $\Gamma$ with several parameters for fixed $N_\mathrm{H}$ (cols.\ 4 and 5) and mixed $N_\mathrm{H}$ (cols.\ 6 and 7). The values in brackets are for photon counts of 50. Spearman correlation coefficient and p-value of no-correlation are given for all the correlations.}
    \label{tab:corr}
\end{table}

\begin{table}
\resizebox{1\linewidth}{!}{
    \centering
\begin{tabular}{l l l l l l} \hline  \hline
Col. & y    &  x         & slope     & intercept
    & ${{\sigma}^2}_{\mathrm{int}}$ \\ 
(1)  & (2)  & (3)        & (4)   & (5)    & (6)  \\ \hline
1   & $\Gamma$ & $\log L_{5100}$   & 0.22 $\pm$ 0.04 (0.42 $\pm$ 0.06)   & -7.22 $\pm$ 1.72(-15.82 $\pm$ 2.47) & 0.14 $\pm$ 0.02(0.1 $\pm$ 0.02)   \\
2   & $\Gamma$ & $R_{4570}$   & 0.20 $\pm$ 0.06 (0.35 $\pm$ 0.10)   & 2.41 $\pm$ 0.05(2.36 $\pm$ 0.09)     & 0.15 $\pm$ 0.02(0.14 $\pm$ 0.03)   \\
3   & $\Gamma$ & $\log M_{BH}$     & 0.13 $\pm$ 0.06 (0.37 $\pm$ 0.10)   & 1.58 $\pm$ 0.43(0.02 $\pm$ 0.75) & 0.16 $\pm$ 0.02(0.14 $\pm$ 0.02) \\
4   & $\Gamma$ & $\log R_{EDD}$    & 0.68 $\pm$ 0.08 (0.91 $\pm$ 0.10)   & 2.89 $\pm$ 0.04(3.13 $\pm$ 0.06) & 0.12 $\pm$ 0.02(0.07 $\pm$ 0.02)  \\
6   & $\Gamma$ & v$_{out}$/1e3    & 1.51 $\pm$ 0.34 (2.08 $\pm$ 0.55)   & 2.20 $\pm$ 0.09 (2.14 $\pm$ 0.15) & 0.15 $\pm$ 0.02 (0.15 $\pm$ 0.04) \\
7   & $\Gamma$ & AI[OIII]     & -0.45 $\pm$ 0.21 (-0.91 $\pm$ 0.38)  & 2.50 $\pm$ 0.05 (2.5 $\pm$ 0.09)  & 0.16 $\pm$ 0.03 (0.18 $\pm$ 0.04) \\
8   & $\Gamma$ & AI(H$\beta$) & -0.74 $\pm$ 0.32 (-1.75 $\pm$ 0.66)  & 2.57 $\pm$ 0.03 (2.65 $\pm$ 0.05) & 0.17 $\pm$ 0.03 (0.17 $\pm$ 0.04) \\
 \hline
 \hline
    \end{tabular} }
    \caption{Best fit relation using LINMIX code for fixed $N_\mathrm{H}$ and photon counts larger than 10. Col.\ 2 is the y-variable, col.\ 3 is the x-variable, col.\ 4 is the slope, col.\ 5 is the intercept and col.\ 6 is the variance of the intrinsic scatter.}
    \label{tab:linmix}
\end{table}

\subsection{Correlation between outflow indicators and photon index}\label{sec:correlation-study}


\cite{2018Rakshit}
studied
[O III $\lambda$5007]
kinematics of a large number of Type 1 AGN using SDSS spectra and found the kinematic signature of the nonvirial motion in the majority of AGN. The
[O III $\lambda$5007]
velocity shift and dispersion are found to be driven by the Eddington ratio suggesting that the radiation pressure or wind from the accretion disc could drive these outflows to kpc scale. \cite{2022Jha} found strong blue asymmetry in H$\beta$ line indicating strong outflows NLS1 compared to the BLS1 which are dominated by infall.

To investigate any link between the ionised outflows at X-rays generated in the corona close to the black hole as traced by the soft X-ray photon index and the large-scale outflow that is seen in
[O III $\lambda$5007]
and H$\beta$ kinematics, we have plotted in Fig.~\ref{gamma_AI_corr} the correlation between $\Gamma$ with various outflow indicators. Here, we have excluded objects with a low velocity shift of $|V[O III]| <10~$km/s. In particular, we found that a majority of the sources shows blue shifted (negative velocity) [O III $\lambda$5007] lines with velocities in the range -100 km/s to -300 km/s. Additionally, those high blue shifted objects also have larger velocity dispersion up to $\sim$400 km/s. Therefore, the velocity-dispersion diagram (VVD) diagram
(upper left panel of Fig.~\ref{gamma_AI_corr})
shows clear outflow signatures where high photon indices have higher blue shifts and dispersion. We also plotted the outflow velocity which is represented as $V_{\mathrm{out}} =\sqrt({V[OIII]}^2 +{\sigma_{[OIII]}}^2)$. We found $V_{\mathrm{out}}$ positively correlate with the $\Gamma$. The asymmetry indices calculated from the [O III $\lambda$5007] and broad H$\beta$ model profile are shown in the bottom panel of Fig.~\ref{gamma_AI_corr}.

\noindent

The majority of the objects have negative asymmetry in
[O III $\lambda$5007]
which is a signature of outflow.
A similar trend is also seen in the case of H$\beta$ asymmetry.

\subsection{Outflow energetics}
We calculated the energetics of the gas outflows e.g., mass accretion rate, energy injection, and power output based on the simple bi-conical outflow and case B recombination
\citep{2010Crenshaw,2016BaeWoo,2018Rakshit}.
Using the [OIII $\lambda$5007] luminosity, we first estimated the mass of the ionised gas following \cite{2011NesvadbaB}

\begin{equation}
M_\mathrm{gas} = 0.4 \times 10^8 M_{\odot} \times (L_{\mathrm{[O\, III],43}}) (100 \, \mathrm{cm^{-3}}/n_e)
\label{eq:mass}
\end{equation}
here $n_e$ is the electron density, which depends on several factors e.g., gas temperature, ionisation mechanism, and geometry of the Narrow-Line Region (NLR). We took the median value of $n_e=272$ from
\cite{2018Rakshit}
and calculated the gas mass. We calculated the size of the gas outflow ($R_{\mathrm{out}}$) based on the empirical relation based on the integral field spectroscopy by \citet{2018ApJ...864..124K}.

\begin{equation}
\log R_{\mathrm{out}} (\mathrm{kpc}) = (0.28\pm0.03) \times \log L_{\mathrm{[O\,III]}} - (11.27\pm 1.46).
\end{equation}
which ranges between 0.6-7.2 kpc with a median of 2 kpc. At the median radius, the  mass outflow rate ($\dot{M}_{\mathrm{out}}$), energy injection rates ($\dot{E}_{\mathrm{out}}$) and momentum flux ($\dot{P}_{\mathrm{out}}$) can be calculated as $\dot{M}_{\mathrm{out}}  =   3 M_{\mathrm{gas}} \frac{v_{\mathrm{out}}}{R_{\mathrm{out}}}$, $\dot{E}_{\mathrm{out}} =    \frac{1}{2} \dot{M}_{\mathrm{out}} v^2_{\mathrm{out}}$, and $\dot{P}_{\mathrm{out}} = \dot{M}_{\mathrm{out}} v_{\mathrm{out}}$. Here, $v_{\mathrm{out}}= \sqrt(v[OIII]^2 + \sigma[OIII]^2)$ is the outflow velocity.

\begin{figure}
\includegraphics[width=\columnwidth]{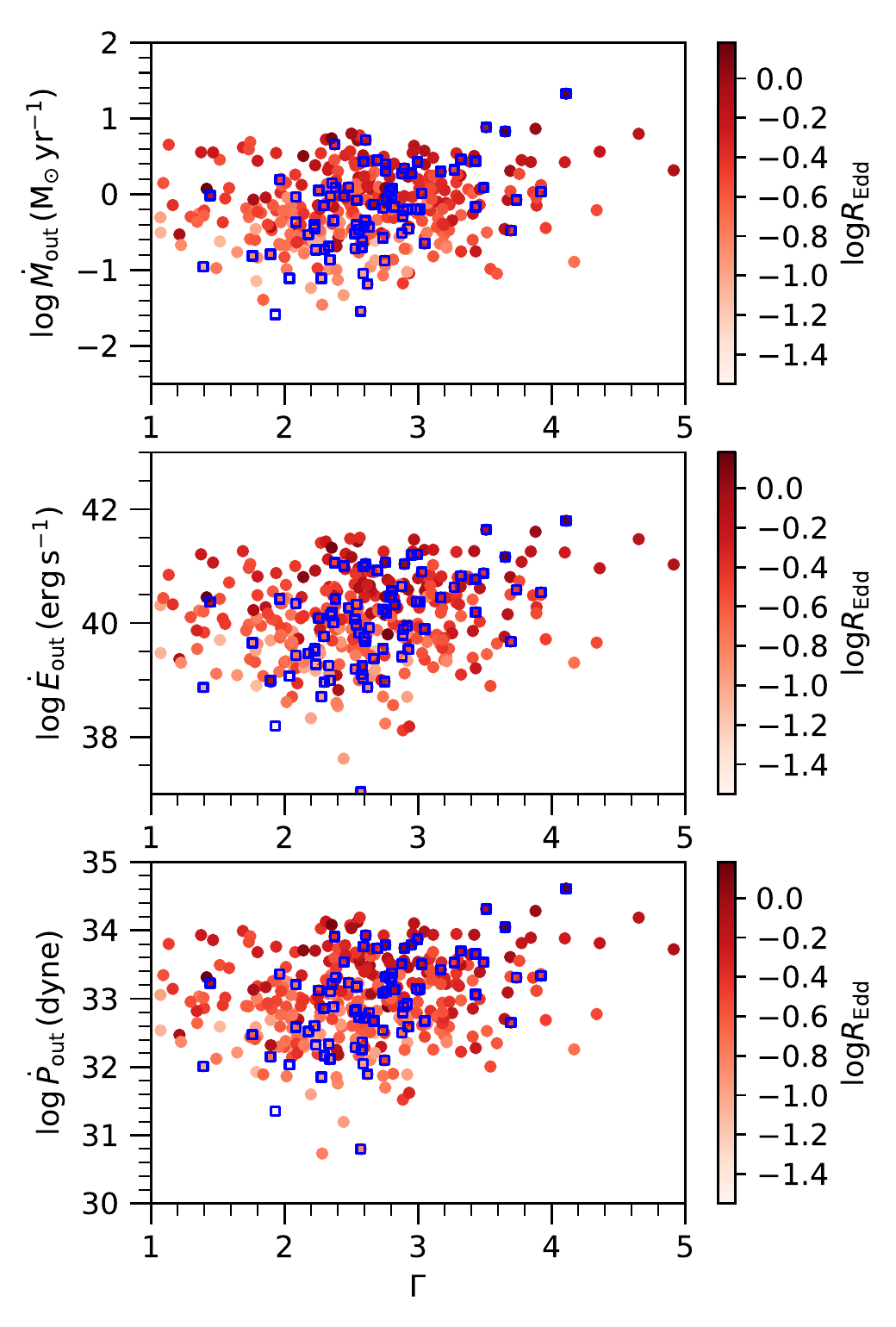}
\caption{Mass outflow rate (top), energy injection rate (middle), and power output (bottom) are shown as a function of $\Gamma$ with colour coded by Eddington ratio for fixed $N_\mathrm{H}$ for sources with photon counts larger than 10 (red-dot) and 50 (blue-dot).}
\label{outflow}
\end{figure}

In Fig.~\ref{outflow}, we show the outflow energetics for the NLS1 in our sample as a function of photon index colour-coded by the Eddington ratio. The mass outflow rate ranges between $0.02-21\, M_{\odot}\, \mathrm{yr^{-1}}$ with a median at $\sim 0.66 \, M_{\odot}\, \mathrm{yr^{-1}}$, the energy injection rate is found to be $6\times10^{36} - 6 \times 10^{41}$ erg/s with a median of $1\times10^{40}$, and the momentum flux ranges between $5\times10^{30} - 4\times 10^{34}$ dyne with a median of $1\times10^{33}$ dyne.
\cite{2018Rakshit}
calculated the outflow energetics for a large number of low-z type 1 AGN and found the median values of $\dot{M}_{\mathrm{out}}$, $\dot{E}_{\mathrm{out}}$ and $\dot{P}_{\mathrm{out}}$ of $0.48 M_{\odot} yr^{-1}$, $1\times10^{40}$ erg s$^{-1}$ and $8\times10^{32}$. Our measurements of outflow energetics of NLS1 are consistent with the values estimated for type 1 AGN.

Outflow energetics are found to be positively correlated with the photon index suggesting $\dot{M}_{\mathrm{out}}$, $\dot{E}_{\mathrm{out}}$ and $\dot{P}_{\mathrm{out}}$ is larger with steep spectra. In fact, the outflow energetics are positively correlated with the Eddington ratio. The plot further suggests that steep spectral sources are accreting at high Eddington rates and have larger mass outflow rate, energy injection, and momentum flux.

\section{Timing analysis of the DR12 NLS1 content}\label{sec:timing}

We have created light curves for all NLS1s detected in the first eROSITA all-sky survey scan (eRASS1). In addition, we have compared these eRASS1 light curves with light curves detected in the second and third scans. We have performed basic variability tests for the individual light curves.
The survey data does not have sufficient exposure time to make definitive claims about variability, but some sources can be considered as variable candidates.
We found 39 objects of the DR12 catalog, which matched with eRASS1 and
exhibit significance values below 3\,$\sigma$ for at least one of the three surveys.
In the Appendix, we list in
Tab.~\ref{tab:variablesources_amplmax_DR12}.
objects which show indications for X-ray variability. If one of the three surveys is not listed in the table, there is no sign of variability for this certain object in the certain survey. Only the surveys, which show variability, are presented.

\section{Discussions}\label{sec:discussion}

AGN outflows detected in the NLR and the potential connections to the feedback processes related to accretion physics have been proposed in several papers. Below we discuss some previous basic findings and compare them to the eROSITA analysis presented in this paper.


\subsection{Relations between the hard 2-10 keV photon index and the [O III \texorpdfstring{$\lambda$}{}5007] line profile}\label{sec:relations}

A sample of 47 type 1 AGN extracted from the XMMi/SDSS-DR7 catalog was analysed by \cite{2016Wang}.
The sample contains both NLS1s and BLS1s. A correlation between the
[O III $\lambda$5007]
velocity shift and the velocity dispersion (their Fig.~7) was found, which is interpreted as wind/radiation-driven outflow.
The authors argue that the hard 2-10\,keV photon index scales with the Eddington ratio which is interpreted as a moderate sign for general AGN outflows. It is argued that a cool-downed corona results in steeper hard photon indices. A general discussion on the accretion rate corona temperature relation is given in
\cite{2011Boller}. Interestingly, Fe K-edged absorption features of O VII ions at 0.74\,keV (their Fig.~1) have been detected, for 3 out of the 47 objects.

\subsection{Gas outflows from large samples based on [O III \texorpdfstring{$\lambda$}{}5007] kinematics}\label{sec:outflow_kinematics}

One of the recent analyses on a large sample of about 5000 type-1 AGN was performed by \cite{2018Rakshit}.
The study is based on detailed outflow parameter determinations including velocity shifts and velocity dispersions of the
[O III $\lambda$5007]
line as well as combining both parameters (VVD diagrams). The outflow energetics have been determined by calculating the mass outflow rate, the energy injection rate, and the momentum flux. These parameters have been correlated with the bolometric luminosity and the relation to the accretion rates has been derived. Based on the
[O III $\lambda$5007]
velocity dispersion and VVD extending to higher values for higher luminosities and Eddington ratios, radiation pressure or winds are concluded to cause the outflow.

\subsection{Soft X-ray studies and kpc-scale outflow studies before eROSITA}\label{sec:outflow_rosat_xmm}

\cite{2022Jha} have presented a study of 144 NLS1s and 117 BLS1s based on ROSAT and XMM-Newton observations.
For their NLS1 sample, the fraction of objects showing outflow signatures as negative AI was three times higher than for the BLS1 sample.
In their Fig.~7 they show the correlation of the outflow signatures with respect to the photon index, Fe II strength, Eddington ratios, and nuclear luminosities. The correlations between these parameters have in general higher probability values compared to the studies presented in this work, mainly due to the inclusion of BLS1 objects, enlarging the parameter space.

\subsection{Evidence for outflow from the gravitational radius up to the kpc scale as seen by eROSITA}\label{sec:outflow_erosita}

In this paper, we have used the asymmetry index and the soft X-ray photon index for the first time to study AGN outflows based on eROSITA NLS1s DR12 detections.
The superior soft-energy response of eROSITA allows us unique studies of the relation of the X-ray spectral steepness in the soft X-ray energy band with optical outflow indicators.
In addition, this is  the largest sample of NLS1s where the soft X-ray photon is applied for outflow studies.

A connection between the accretion disk and the outflow in the NLR has been suggested by \cite{2016Wang} based on the hard 2-10\,keV photon index. The authors argue that the hard X-ray photon index is linked to the accretion process. In this scenario, the hard X-ray photon index is produced by inverse Compton scattering of accretion disk photons in the hot about $\rm 10^9$ K corona.

In this paper, we use the soft X-ray photon index from the eROSITA observations to study the launching mechanism for AGN outflows.
Ultra-fast ionised outflows with high covering fractions can potentially cause a moderately steep powerlaw to appear extremely steep due to blue-shifted absorption features of Fe L and lighter elements causing a depression of photons just above 1\,keV. The observed photon index can then reach values well above five as can be seen when simulating data using XSPEC's \texttt{zxipcf}-model. 
The eROSITA soft X-ray photon index is therefore a better tracer for AGN outflows at the innermost gravitational radius scale compared to studies based on the hard X-ray photon index which probes mainly Comptonisation effects.

The correlation between $\Gamma$ and $\rm H\beta$ asymmetry is weak but still present and stronger for photon counts above 50. We note that the geometry and kinematics of the BLR are highly complex. The low-ionisation broad line such as H$\beta$ on average does not show any significant blueshift/redshift with respect to the systemic velocity (as also evident from Figure that majority of objects have AI(H$\beta$) =0) compared to the high ionisation line e.g., CIV, dominated by Keplerian motion. However, high-quality reverberation mapping study shows the presence of inflow and outflow in H$\beta$ of a few AGNs \citep[e.g.,][]{2017ApJ...849..146G}.  Hence, the presence of net radial motion and/or opacity effects could produce small H$\beta$ asymmetry in some AGNs \citep{2012ApJS..201...23E,2016ApJ...831....7S}.
We find strong correlations between the soft X-ray photon index and the Eddington rate, the Fe II strength and with the [O III $\lambda$5007] outflow velocity-, dispersion- and the AI value.

The eROSITA studies in combination with new asymmetry parameter calculations done for the \cite{RS017} sample provides strong evidence for the wind/radiation driven mode of AGN feedback processes in NLS1s.

\noindent

%

%
%

\section{Summary and Outlook}\label{sec:summary}

We provide the first look at NLS1s based on the to-date largest spectroscopically confirmed NLS1 sample \citep{RS017} with the first X-ray all-sky survey scan of the eROSITA mission.
The spectra were modelled, and the physically expected correlations between the spectral steepness and optical continuum and emission line parameters could be reproduced and extended in the parameter spaces.

A single redshifted powerlaw model was adopted to model the entire spectrum.
A combined black-body and powerlaw model to account for the soft X-ray excess emission and the coronal emission was tested, however the fits performed were deemed unreliable \citep{Gruenwald2022}.
The mean photon index is steep with about 2.8 as expected, however, an ultra-soft photon index tail has been found to skew the distribution to values as high as 10.
This will be investigated further.

The NLS1 sample of \cite{RS017} was extended by re-fitting all spectra to determine the line asymmetry and outflow indicators. The dependency of the photon index with AGN parameters, correlations between outflow indicators and the photon index, and the outflow energetics have been determined.
We note that the eROSITA soft X-ray photon index correlates with the large-scale [O III] outflow.

We argue that ionised outflows from the gravitational radius scale up to the NLR region
provide the best physical explanation for the
eROSITA NLS1 SDSS DR12 data.



\begin{acknowledgements}

We thank the anonymous referee for their helpful comments and suggestions which helped to improve the quality of the paper. SR acknowledges the partial support of SRG-SERB, DST, New Delhi through grant no. SRG/2021/001334. 
TB is grateful to the XMM-Newton Project Scientist Dr. Norbert Schartel for accepting DDT time for 4 of the objects. The results will be reported elsewhere.
This work is based on data from eROSITA, the soft X-ray instrument aboard SRG, a joint Russian-German science mission supported by the Russian Space Agency (Roskosmos), in the interests of the Russian Academy of Sciences represented by its Space Research Institute (IKI), and the Deutsches Zentrum für Luft- und Raumfahrt (DLR). The SRG spacecraft was built by Lavochkin Association (NPOL) and its subcontractors, and is operated by NPOL with support from the Max Planck Institute for Extraterrestrial Physics (MPE).
The development and construction of the eROSITA X-ray instrument was led by MPE, with contributions from the Dr. Karl Remeis Observatory Bamberg \& ECAP (FAU Erlangen-Nuernberg), the University of Hamburg Observatory, the Leibniz Institute for Astrophysics Potsdam (AIP), and the Institute for Astronomy and Astrophysics of the University of Tübingen, with the support of DLR and the Max Planck Society. The Argelander Institute for Astronomy of the University of Bonn and the Ludwig Maximilians Universität Munich also participated in the science preparation for eROSITA.
The eROSITA data shown here were processed using the eSASS/NRTA software system developed by the German eROSITA consortium.

\end{acknowledgements}

\bibliographystyle{aa}
\bibliography{main.bbl}

\begin{thebibliography}{61}
\expandafter\ifx\csname natexlab\endcsname\relax\def\natexlab#1{#1}\fi

\bibitem[{{Anderson} {et~al.}(2007){Anderson}, {Margon}, {Voges}, {Plotkin},
  {Syphers}, {Haggard}, {Collinge}, {Meyer}, {Strauss}, {Ag{\"u}eros}, {Hall},
  {Homer}, {Ivezi{\'c}}, {Richards}, {Richmond}, {Schneider}, {Stinson},
  {Vanden Berk}, \& {York}}]{2007Anderson}
{Anderson}, S.~F., {Margon}, B., {Voges}, W., {et~al.} 2007, \aj, 133, 313

\bibitem[{{Arav} {et~al.}(2013){Arav}, {Borguet}, {Chamberlain}, {Edmonds}, \&
  {Danforth}}]{2013Arav}
{Arav}, N., {Borguet}, B., {Chamberlain}, C., {Edmonds}, D., \& {Danforth}, C.
  2013, \mnras, 436, 3286

\bibitem[{{Bae} \& {Woo}(2016)}]{2016BaeWoo}
{Bae}, H.-J. \& {Woo}, J.-H. 2016, \apj, 828, 97

\bibitem[{{Bentz} {et~al.}(2013){Bentz}, {Denney}, {Grier}, {Barth},
  {Peterson}, {Vestergaard}, {Bennert}, {Canalizo}, {De Rosa}, {Filippenko},
  {Gates}, {Greene}, {Li}, {Malkan}, {Pogge}, {Stern}, {Treu}, \&
  {Woo}}]{2013Bentz}
{Bentz}, M.~C., {Denney}, K.~D., {Grier}, C.~J., {et~al.} 2013, \apj, 767, 149

\bibitem[{{Boller} {et~al.}(1996){Boller}, {Brandt}, \& {Fink}}]{1996Boller}
{Boller}, T., {Brandt}, W.~N., \& {Fink}, H. 1996, \aap, 305, 53

\bibitem[{{Boller} {et~al.}(2016){Boller}, {Freyberg}, {Tr{\"u}mper}, {Haberl},
  {Voges}, \& {Nandra}}]{Boller2016}
{Boller}, T., {Freyberg}, M.~J., {Tr{\"u}mper}, J., {et~al.} 2016, \aap, 588,
  A103

\bibitem[{{Boller} {et~al.}(2011){Boller}, {Schady}, \&
  {Heftrich}}]{2011Boller}
{Boller}, T., {Schady}, P., \& {Heftrich}, T. 2011, \apjl, 731, L16

\bibitem[{{Boller} {et~al.}(2022){Boller}, {Schmitt}, {Buchner}, {Freyberg},
  {Georgakakis}, {Liu}, {Robrade}, {Merloni}, {Nandra}, {Malyali}, {Krumpe},
  {Salvato}, \& {Dwelly}}]{2022Boller}
{Boller}, T., {Schmitt}, J.~H.~M.~M., {Buchner}, J., {et~al.} 2022, \aap, 661,
  A8

\bibitem[{{Boroson} \& {Green}(1992)}]{1992Boroson}
{Boroson}, T.~A. \& {Green}, R.~F. 1992, \apjs, 80, 109

\bibitem[{{Buchner} {et~al.}(2014){Buchner}, {Georgakakis}, {Nandra}, {Hsu},
  {Rangel}, {Brightman}, {Merloni}, {Salvato}, {Donley}, \&
  {Kocevski}}]{Buchner2014}
{Buchner}, J., {Georgakakis}, A., {Nandra}, K., {et~al.} 2014, A\&A, 564, A125

\bibitem[{{Crenshaw} {et~al.}(2003){Crenshaw}, {Kraemer}, \&
  {George}}]{2003ARA&A..41..117Crenshaw}
{Crenshaw}, D.~M., {Kraemer}, S.~B., \& {George}, I.~M. 2003, \araa, 41, 117

\bibitem[{{Crenshaw} {et~al.}(2010){Crenshaw}, {Schmitt}, {Kraemer},
  {Mushotzky}, \& {Dunn}}]{2010Crenshaw}
{Crenshaw}, D.~M., {Schmitt}, H.~R., {Kraemer}, S.~B., {Mushotzky}, R.~F., \&
  {Dunn}, J.~P. 2010, \apj, 708, 419

\bibitem[{{Eracleous} {et~al.}(2012){Eracleous}, {Boroson}, {Halpern}, \&
  {Liu}}]{2012ApJS..201...23E}
{Eracleous}, M., {Boroson}, T.~A., {Halpern}, J.~P., \& {Liu}, J. 2012, \apjs,
  201, 23

\bibitem[{{Fabian}(2012)}]{2012Fabian}
{Fabian}, A.~C. 2012, \araa, 50, 455

\bibitem[{{Fitzpatrick}(1999)}]{1999PASP..111...63F}
{Fitzpatrick}, E.~L. 1999, \pasp, 111, 63

\bibitem[{{Goodrich}(1989)}]{Goodrich1989}
{Goodrich}, R.~W. 1989, ApJ, 342, 224

\bibitem[{{Grier} {et~al.}(2017){Grier}, {Pancoast}, {Barth}, {Fausnaugh},
  {Brewer}, {Treu}, \& {Peterson}}]{2017ApJ...849..146G}
{Grier}, C.~J., {Pancoast}, A., {Barth}, A.~J., {et~al.} 2017, \apj, 849, 146

\bibitem[{{Grünwald}(2022)}]{Gruenwald2022}
{Grünwald}, G. 2022, {Master} thesis, Goethe University of Frankfurt, Germany

\bibitem[{{Guo} {et~al.}(2018){Guo}, {Shen}, \& {Wang}}]{2018ascl.soft09008G}
{Guo}, H., {Shen}, Y., \& {Wang}, S. 2018, {PyQSOFit: Python code to fit the
  spectrum of quasars}, Astrophysics Source Code Library

\bibitem[{{Heckman} {et~al.}(1981){Heckman}, {Miley}, {van Breugel}, \&
  {Butcher}}]{1981Heckman}
{Heckman}, T.~M., {Miley}, G.~K., {van Breugel}, W.~J.~M., \& {Butcher}, H.~R.
  1981, \apj, 247, 403

\bibitem[{Jha {et~al.}(2021)Jha, Chand, Ojha, Omar, \& Rastogi}]{2022Jha}
Jha, V.~K., Chand, H., Ojha, V., Omar, A., \& Rastogi, S. 2021, \mnras, 510,
  4379

\bibitem[{{Kang} \& {Woo}(2018)}]{2018ApJ...864..124K}
{Kang}, D. \& {Woo}, J.-H. 2018, \apj, 864, 124

\bibitem[{{Kelly}(2007)}]{2007ApJ...665.1489K}
{Kelly}, B.~C. 2007, \apj, 665, 1489

\bibitem[{{King} \& {Pounds}(2015)}]{2015King}
{King}, A. \& {Pounds}, K. 2015, \araa, 53, 115

\bibitem[{{King} \& {Pounds}(2003)}]{2003King}
{King}, A.~R. \& {Pounds}, K.~A. 2003, \mnras, 345, 657

\bibitem[{{Kormendy} \& {Ho}(2013)}]{2013Kormendy}
{Kormendy}, J. \& {Ho}, L.~C. 2013, \araa, 51, 511

\bibitem[{{Liu} {et~al.}(2021){Liu}, {Buchner}, {Nandra}, {Merloni}, {Dwelly},
  {Sanders}, {Salvato}, {Arcodia}, {Brusa}, {Wolf}, {Georgakakis}, {Boller},
  {Krumpe}, {Lamer}, {Waddell}, {Urrutia}, {Schwope}, {Robrade}, {Wilms},
  {Dauser}, {Comparat}, {Toba}, {Ichikawa}, {Iwasawa}, {Shen}, \& {Ibarra
  Medel}}]{Liu2021}
{Liu}, T., {Buchner}, J., {Nandra}, K., {et~al.} 2021, arXiv e-prints,
  arXiv:2106.14522

\bibitem[{{Liu} {et~al.}(2016){Liu}, {Merloni}, {Georgakakis}, {Menzel},
  {Buchner}, {Nandra}, {Salvato}, {Shen}, {Brusa}, \&
  {Streblyanska}}]{LiuZhu2016}
{Liu}, Z., {Merloni}, A., {Georgakakis}, A., {et~al.} 2016, \mnras, 459, 1602

\bibitem[{{Marziani} {et~al.}(1996){Marziani}, {Sulentic}, {Dultzin-Hacyan},
  {Calvani}, \& {Moles}}]{1996Marziani}
{Marziani}, P., {Sulentic}, J.~W., {Dultzin-Hacyan}, D., {Calvani}, M., \&
  {Moles}, M. 1996, \apjs, 104, 37

\bibitem[{{Merloni} \& {Consortium}(2023)}]{Merloni2023}
{Merloni}, A. \& {Consortium}. 2023, in prep.

\bibitem[{{Merloni} {et~al.}(2012){Merloni}, {Predehl}, {Becker},
  {B{\"o}hringer}, {Boller}, {Brunner}, {Brusa}, {Dennerl}, {Freyberg},
  {Friedrich}, {Georgakakis}, {Haberl}, {Hasinger}, {Meidinger}, {Mohr},
  {Nandra}, {Rau}, {Reiprich}, {Robrade}, {Salvato}, {Santangelo}, {Sasaki},
  {Schwope}, {Wilms}, \& {German eROSITA Consortium}}]{2012Merloni}
{Merloni}, A., {Predehl}, P., {Becker}, W., {et~al.} 2012, arXiv e-prints,
  arXiv:1209.3114

\bibitem[{{Nesvadba} {et~al.}(2011){Nesvadba}, {Polletta}, {Lehnert},
  {Bergeron}, {De Breuck}, {Lagache}, \& {Omont}}]{2011NesvadbaB}
{Nesvadba}, N.~P.~H., {Polletta}, M., {Lehnert}, M.~D., {et~al.} 2011, \mnras,
  415, 2359

\bibitem[{{Ojha} {et~al.}(2020){Ojha}, {Chand}, {Dewangan}, \&
  {Rakshit}}]{Ojha2020}
{Ojha}, V., {Chand}, H., {Dewangan}, G.~C., \& {Rakshit}, S. 2020, \apj, 896,
  95

\bibitem[{{Osterbrock} \& {Pogge}(1985)}]{OP1985}
{Osterbrock}, D.~E. \& {Pogge}, R.~W. 1985, ApJ, 297, 166

\bibitem[{Page {et~al.}(2012)Page, Brindle, Talavera, Still, Rosen, Yershov,
  Ziaeepour, Mason, Cropper, Breeveld, Loiseau, Mignani, Smith, \&
  Murdin}]{Page2012}
Page, M.~J., Brindle, C., Talavera, A., {et~al.} 2012, Monthly Notices of the
  Royal Astronomical Society, 426, 903

\bibitem[{{Parker} {et~al.}(2021){Parker}, {Alston}, {H{\"a}rer}, {Igo},
  {Joyce}, {Buisson}, {Chainakun}, {Fabian}, {Jiang}, {Kosec}, {Matzeu},
  {Pinto}, {Xu}, \& {Zaidouni}}]{2021Parker}
{Parker}, M.~L., {Alston}, W.~N., {H{\"a}rer}, L., {et~al.} 2021, \mnras, 508,
  1798

\bibitem[{{Pounds} {et~al.}(2003{\natexlab{a}}){Pounds}, {King}, {Page}, \&
  {O'Brien}}]{2003Poundsa}
{Pounds}, K.~A., {King}, A.~R., {Page}, K.~L., \& {O'Brien}, P.~T.
  2003{\natexlab{a}}, \mnras, 346, 1025

\bibitem[{{Pounds} {et~al.}(2003{\natexlab{b}}){Pounds}, {Reeves}, {King},
  {Page}, {O'Brien}, \& {Turner}}]{2003Poundsb}
{Pounds}, K.~A., {Reeves}, J.~N., {King}, A.~R., {et~al.} 2003{\natexlab{b}},
  \mnras, 345, 705

\bibitem[{{Pounds} {et~al.}(2003{\natexlab{c}}){Pounds}, {Reeves}, {Page},
  {Wynn}, \& {O'Brien}}]{2003Poundsc}
{Pounds}, K.~A., {Reeves}, J.~N., {Page}, K.~L., {Wynn}, G.~A., \& {O'Brien},
  P.~T. 2003{\natexlab{c}}, \mnras, 342, 1147

\bibitem[{{Predehl} {et~al.}(2021){Predehl}, {Andritschke}, {Arefiev},
  {Babyshkin}, {Batanov}, {Becker}, {B{\"o}hringer}, {Bogomolov}, {Boller},
  {Borm}, {Bornemann}, {Br{\"a}uninger}, {Br{\"u}ggen}, {Brunner}, {Brusa},
  {Bulbul}, {Buntov}, {Burwitz}, {Burkert}, {Clerc}, {Churazov}, {Coutinho},
  {Dauser}, {Dennerl}, {Doroshenko}, {Eder}, {Emberger}, {Eraerds},
  {Finoguenov}, {Freyberg}, {Friedrich}, {Friedrich}, {F{\"u}rmetz},
  {Georgakakis}, {Gilfanov}, {Granato}, {Grossberger}, {Gueguen}, {Gureev},
  {Haberl}, {H{\"a}lker}, {Hartner}, {Hasinger}, {Huber}, {Ji}, {Kienlin},
  {Kink}, {Korotkov}, {Kreykenbohm}, {Lamer}, {Lomakin}, {Lapshov}, {Liu},
  {Maitra}, {Meidinger}, {Menz}, {Merloni}, {Mernik}, {Mican}, {Mohr},
  {M{\"u}ller}, {Nandra}, {Nazarov}, {Pacaud}, {Pavlinsky}, {Perinati},
  {Pfeffermann}, {Pietschner}, {Ramos-Ceja}, {Rau}, {Reiffers}, {Reiprich},
  {Robrade}, {Salvato}, {Sanders}, {Santangelo}, {Sasaki}, {Scheuerle},
  {Schmid}, {Schmitt}, {Schwope}, {Shirshakov}, {Steinmetz}, {Stewart},
  {Str{\"u}der}, {Sunyaev}, {Tenzer}, {Tiedemann}, {Tr{\"u}mper}, {Voron},
  {Weber}, {Wilms}, \& {Yaroshenko}}]{2021Predehl}
{Predehl}, P., {Andritschke}, R., {Arefiev}, V., {et~al.} 2021, \aap, 647, A1

\bibitem[{{Rakshit} {et~al.}(2017){Rakshit}, {Stalin}, {Chand}, \&
  {Zhang}}]{RS017}
{Rakshit}, S., {Stalin}, C.~S., {Chand}, H., \& {Zhang}, X.-G. 2017, \apjs,
  229, 39

\bibitem[{Rakshit {et~al.}(2017)Rakshit, Stalin, Chand, \& Zhang}]{2017Rakshit}
Rakshit, S., Stalin, C.~S., Chand, H., \& Zhang, X.-G. 2017, The Astrophysical
  Journal Supplement Series, 229, 39

\bibitem[{{Rakshit} {et~al.}(2020){Rakshit}, {Stalin}, \&
  {Kotilainen}}]{2020ApJS..249...17R}
{Rakshit}, S., {Stalin}, C.~S., \& {Kotilainen}, J. 2020, \apjs, 249, 17

\bibitem[{{Rakshit} \& {Woo}(2018)}]{2018Rakshit}
{Rakshit}, S. \& {Woo}, J.-H. 2018, \apj, 865, 5

\bibitem[{{Reeves} {et~al.}(2009){Reeves}, {O'Brien}, {Braito}, {Behar},
  {Miller}, {Turner}, {Fabian}, {Kaspi}, {Mushotzky}, \&
  {Ward}}]{2009ApJ...701..493Reeves}
{Reeves}, J.~N., {O'Brien}, P.~T., {Braito}, V., {et~al.} 2009, \apj, 701, 493

\bibitem[{{Reeves} {et~al.}(2003){Reeves}, {O'Brien}, \& {Ward}}]{2003Reeves}
{Reeves}, J.~N., {O'Brien}, P.~T., \& {Ward}, M.~J. 2003, \apjl, 593, L65

\bibitem[{{Salvato} {et~al.}(2018){Salvato}, {Buchner}, {Budav{\'a}ri},
  {Dwelly}, {Merloni}, {Brusa}, {Rau}, {Fotopoulou}, \& {Nandra}}]{Salvato2018}
{Salvato}, M., {Buchner}, J., {Budav{\'a}ri}, T., {et~al.} 2018, \mnras, 473,
  4937

\bibitem[{{Schlegel} {et~al.}(1998){Schlegel}, {Finkbeiner}, \&
  {Davis}}]{1998ApJ...500..525S}
{Schlegel}, D.~J., {Finkbeiner}, D.~P., \& {Davis}, M. 1998, \apj, 500, 525

\bibitem[{{Shen} {et~al.}(2016){Shen}, {Brandt}, {Richards}, {Denney},
  {Greene}, {Grier}, {Ho}, {Peterson}, {Petitjean}, {Schneider}, {Tao}, \&
  {Trump}}]{2016ApJ...831....7S}
{Shen}, Y., {Brandt}, W.~N., {Richards}, G.~T., {et~al.} 2016, \apj, 831, 7

\bibitem[{{Shen} {et~al.}(2019){Shen}, {Hall}, {Horne}, {Zhu}, {McGreer},
  {Simm}, {Trump}, {Kinemuchi}, {Brandt}, {Green}, {Grier}, {Guo}, {Ho},
  {Homayouni}, {Jiang}, {I-Hsiu Li}, {Morganson}, {Petitjean}, {Richards},
  {Schneider}, {Starkey}, {Wang}, {Chambers}, {Kaiser}, {Kudritzki}, {Magnier},
  \& {Waters}}]{2019ApJS..241...34S}
{Shen}, Y., {Hall}, P.~B., {Horne}, K., {et~al.} 2019, \apjs, 241, 34

\bibitem[{{Simmonds} {et~al.}(2018){Simmonds}, {Buchner}, {Salvato}, {Hsu}, \&
  {Bauer}}]{Simmonds2018}
{Simmonds}, C., {Buchner}, J., {Salvato}, M., {Hsu}, L.~T., \& {Bauer}, F.~E.
  2018, \aap, 618, A66

\bibitem[{{Tombesi} {et~al.}(2010){Tombesi}, {Cappi}, {Reeves}, {Palumbo},
  {Yaqoob}, {Braito}, \& {Dadina}}]{2010Tombesi}
{Tombesi}, F., {Cappi}, M., {Reeves}, J.~N., {et~al.} 2010, \aap, 521, A57

\bibitem[{{Tombesi} {et~al.}(2015){Tombesi}, {Mel{\'e}ndez}, {Veilleux},
  {Reeves}, {Gonz{\'a}lez-Alfonso}, \& {Reynolds}}]{2015Natur.519..436Tombesi}
{Tombesi}, F., {Mel{\'e}ndez}, M., {Veilleux}, S., {et~al.} 2015, \nat, 519,
  436

\bibitem[{{Verner} {et~al.}(1996){Verner}, {Ferland}, {Korista}, \&
  {Yakovlev}}]{Verner1996}
{Verner}, D.~A., {Ferland}, G.~J., {Korista}, K.~T., \& {Yakovlev}, D.~G. 1996,
  ApJ, 465, 487

\bibitem[{{Wang} {et~al.}(2016){Wang}, {Xu}, \& {Wei}}]{2016Wang}
{Wang}, J., {Xu}, D.~W., \& {Wei}, J.~Y. 2016, \aj, 151, 81

\bibitem[{{Webb} {et~al.}(2020){Webb}, {Coriat}, {Traulsen}, {Ballet}, {Motch},
  {Carrera}, {Koliopanos}, {Authier}, {de la Calle}, {Ceballos}, {Colomo},
  {Chuard}, {Freyberg}, {Garcia}, {Kolehmainen}, {Lamer}, {Lin}, {Maggi},
  {Michel}, {Page}, {Page}, {Perea-Calderon}, {Pineau}, {Rodriguez}, {Rosen},
  {Santos Lleo}, {Saxton}, {Schwope}, {Tom{\'a}s}, {Watson}, \&
  {Zakardjian}}]{Webb2020}
{Webb}, N.~A., {Coriat}, M., {Traulsen}, I., {et~al.} 2020, \aap, 641, A136

\bibitem[{{Wilms} {et~al.}(2000){Wilms}, {Allen}, \& {McCray}}]{Wilms2000}
{Wilms}, J., {Allen}, A., \& {McCray}, R. 2000, ApJ, 542, 914

\bibitem[{{Wolf} {et~al.}(2020){Wolf}, {Salvato}, {Coffey}, {Merloni},
  {Buchner}, {Arcodia}, {Baron}, {Carrera}, {Comparat}, {Schneider}, \&
  {Nandra}}]{2020Wolf}
{Wolf}, J., {Salvato}, M., {Coffey}, D., {et~al.} 2020, \mnras, 492, 3580

\bibitem[{{Woo} {et~al.}(2018){Woo}, {Le}, {Karouzos}, {Park}, {Park},
  {Malkan}, {Treu}, \& {Bennert}}]{2018Woo}
{Woo}, J.-H., {Le}, H. A.~N., {Karouzos}, M., {et~al.} 2018, \apj, 859, 138

\bibitem[{{Woo} {et~al.}(2015){Woo}, {Yoon}, {Park}, {Park}, \&
  {Kim}}]{2015Woo}
{Woo}, J.-H., {Yoon}, Y., {Park}, S., {Park}, D., \& {Kim}, S.~C. 2015, \apj,
  801, 38

\bibitem[{{Yip} {et~al.}(2004){Yip}, {Connolly}, {Vanden Berk}, {Ma},
  {Frieman}, {SubbaRao}, {Szalay}, {Richards}, {Hall}, {Schneider}, {Hopkins},
  {Trump}, \& {Brinkmann}}]{2004AJ....128.2603Y}
{Yip}, C.~W., {Connolly}, A.~J., {Vanden Berk}, D.~E., {et~al.} 2004, \aj, 128,
  2603

\end{thebibliography}

\appendix

\section{Detailed investigation of the light curve properties}
\label{sec:AppendixA1}
We have performed standard normalized excess variance (NEV) and maximum amplitude variability (AMPL\_MAX) analysis
as described in \cite{2022Boller}
for the light curves for eROSITA all-sky survey scans eRASS1, eRASS2, and eRASS3, respectively.
In Table \ref{tab:variablesources_amplmax_DR12}
we list only those objects which have significance values in the standard variability tests greater than
zero.

\begin{table*}
    \caption[List of Variable NLS1 Sources DR12 I]{This table shows the variable NLS1 source candidates of SDSS DR12. The objects are sorted decreasingly by the value of \texttt{AMPL\_MAX} and their SDSS ID (plate-mjd-fiber). If one of the three eROSITA surveys is not listed in this table, there is no sign for variability for this certain object in the certain survey. The sources can be identified by their SDSS ID and coordinates. Additional detections are labelled as 2RXS \citep{Boller2016}, XMM \citep{Webb2020}, and XMM-SUSS41 \citep{Page2012}, respectively. {\tt RA} and {\tt DEC} refer to the SDSS coordinates.}
    \label{tab:variablesources_amplmax_DR12}
    \resizebox{\textwidth}{!}{
        \begin{tabular}{lcrrcllll}
            \hline
            \textbf{SDSS ID} & \textbf{eRASS} & \textbf{RA} &\textbf{DEC} & \textbf{AMPL\_MAX} & \textbf{NEV} & \textbf{Object Name} & \textbf{Detection} &  \textbf{Redshift (spectroscopic)}\\
   0522-52024-0173 & 3 & 191.64687  &   2.36910 & 2.16 & 1.23 & LBQS 1244+0238             & 2RXS, XMM                      & 0.04813 $\pm$ 0.00002 \\
   0522-52024-0173 & 2 & 191.64687  &   2.36910 & 1.74 & 0.83 & LBQS 1244+0238             & 2RXS, XMM                      & 0.04813 $\pm$ 0.00002\\
   0522-52024-0173 & 1 & 191.64687  &   2.36910 & 0.45 & 0.00 & LBQS 1244+0238             & 2RXS, XMM                      & 0.04813 $\pm$ 0.00002\\

   1797-54507-0315 & 1 & 197.98398  &   6.81620 & 2.00 & 0.54 & 2MASS J13115615+0648583    & 2RXS                           & 0.12736 $\pm$ 0.00002\\

   2265-53674-0046 & 3 & 119.79352  &  11.86577 & 1.25 & 0.42 & 2MASX J07591042+1151563    & 2RXS                           & 0.05014 $\pm$ 0.00003\\
   2265-53674-0046 & 1 & 119.79352  &  11.86577 & 0.07 & 0.00 & 2MASX J07591042+1151563    & 2RXS                           & 0.05014 $\pm$ 0.00003\\

   1799-53556-0369 & 1 & 200.73097  &   8.16156 & 1.17 & 0.27 & Mrk 1347                   & 2RXS                           & 0.04995 $\pm$ 0.00003\\
   1799-53556-0369 & 3 & 200.73097  &   8.16156 & 0.82 & 0.83 & Mrk 1347                   & 2RXS                           & 0.04995 $\pm$ 0.00003\\
   1799-53556-0369 & 2 & 200.73097  &   8.16156 & 0.54 & 0.07 & Mrk 1347                   & 2RXS                           & 0.04995 $\pm$ 0.00003\\

   1060-52636-0293 & 2 & 117.75593  &  29.23868 & 1.09 & 0.39 & ATO J117.7559+29.2386      & 2RXS, XMM                       & 0.12080 $\pm$ 0.00003\\

   0919-52409-0023 & 2 & 220.29812  &  -2.20977 & 1.06 & 0.31 & LEDA 1098647               & 2RXS                           & 0.08227 $\pm$ 0.00002\\
   0919-52409-0023 & 3 & 220.29812  &  -2.20977 & 0.28 & 0.48 & LEDA 1098647               & 2RXS                           & 0.08227 $\pm$ 0.00002\\

   2346-53734-0045 & 2 & 154.77829  &  23.31049 & 0.74 & 0.04 & 2MASS J10190678+2318378    & 2RXS                           & 0.06456 $\pm$ 0.00003\\

   1768-53442-0547 & 2 & 188.73294  &  15.56561 & 0.64 & 0.00 & IC 3528                    & 2RXS                           & 0.04582 $\pm$ 0.00004\\
   1768-53442-0547 & 1 & 188.73294  &  15.56561 & 0.14 & 0.00 & IC 3528                    & 2RXS                           & 0.04582 $\pm$ 0.00004\\

   1772-53089-0077 & 3 & 197.89859  &  14.41311 & 0.61 & 0.00 & 2MASX J13113562+1424472    & 2RXS                           & 0.11393 $\pm$ 0.00002\\
   1772-53089-0077 & 1 & 197.89859  &  14.41311 & 0.32 & 0.00 & 2MASX J13113562+1424472    & 2RXS                           & 0.11393 $\pm$ 0.00002\\

   2488-54149-0312 & 3 & 164.55616  &  21.20372 & 0.56 & 0.00 & LAMOST J105813.47+211213.4 & \textbf{new eROSITA discovery} & 0.68787 $\pm$ 0.00014\\

   2090-53463-0090 & 3 & 162.48570  &  35.27867 & 0.51 & 0.27 & 2E 2333                    & 2RXS                           & 0.49040 $\pm$ 0.00009\\

   1923-53319-0382 & 3 & 120.25589  &  18.81133 & 0.51 & 0.36 & 2MASS J08010140+1848409    & 2RXS, XMM                      & 0.13954 $\pm$ 0.00001\\
   1923-53319-0382 & 1 & 120.25589  &  18.81133 & 0.10 & 0.00 & 2MASS J08010140+1848409    & 2RXS, XMM                      & 0.13954 $\pm$ 0.00001\\

   0414-51869-0413 & 3 &  51.52814  &   1.24163 & 0.48 & 0.00 & 2MASX J03260674+0114297    & 2RXS                           & 0.12700 $\pm$ 0.00003\\

   0850-52338-0245 & 1 & 197.01267  &   3.85405 & 0.45 & 0.00 & 2MASX J13080303+0351144    & 2RXS                           & 0.07045 $\pm$ 0.00002\\
   0850-52338-0245 & 2 & 197.01267  &   3.85405 & 0.32 & 0.00 & 2MASX J13080303+0351144    & 2RXS                           & 0.07045 $\pm$ 0.00002\\
   0850-52338-0245 & 3 & 197.01267  &   3.85405 & 0.15 & 0.00 & 2MASX J13080303+0351144    & 2RXS                           & 0.07045 $\pm$ 0.00002\\

   3374-54948-0507 & 1 & 189.82842  &  24.52879 & 0.42 & 0.00 & LEDA 3096171               & 2RXS                           & 0.18559 $\pm$ 0.00002\\

   0910-52377-0588 & 1 & 202.92096  &  -1.87012 & 0.39 & 0.00 & 2dFGRS TGN198Z132          & 2RXS, XMM                      & 0.14508 $\pm$ 0.00003\\

   0331-52368-0532 & 2 & 180.61151  &  -1.48757 & 0.34 & 0.00 & LEDA 90159                 & 2RXS, XMM                      & 0.15080 $\pm$ 0.00015\\
   0331-52368-0532 & 1 & 180.61151  &  -1.48757 & 0.14 & 0.00 & LEDA 90159                 & 2RXS, XMM                      & 0.15080 $\pm$ 0.00015\\

   1805-53875-0402 & 2 & 207.73667  &   7.70956 & 0.30 & 0.00 & 2MASS J13505681+0742344    & 2RXS                           & 0.07953 $\pm$ 0.00003\\

   2212-53789-0136 & 2 & 166.52868  &  25.59596 & 0.25 & 0.36 & 2MASS J11060689+2535454    & 2RXS                           & 0.17570 $\pm$ 0.00002\\

   1829-53494-0249 & 1 & 220.67001  &   5.40646 & 0.20 & 0.00 & 2MASX J14424081+0524234    & \textbf{new eROSITA discovery} & 0.11703 $\pm$ 0.00004\\

   0586-52023-0089 & 2 & 219.91783  &   3.09128 & 0.20 & 0.00 & 2MASS J14394028+0305283    & 2RXS                           & 0.26892 $\pm$ 0.00069\\

   0585-52027-0368 & 3 & 216.95121  &   5.03946 & 0.20 & 0.00 & 2MASS J14274829+0502220    & 2RXS                           & 0.10583 $\pm$ 0.00002\\

   2578-54093-0425 & 1 & 141.34329  &  13.09408 & 0.18 & 0.00 & 2MASS J09252237+1305388    & 2RXS                           & 0.07941 $\pm$ 0.00004\\

   2370-53764-0123 & 1 & 147.51519  &  17.15908 & 0.17 & 0.00 & SDSS J095003.64+170932.6   & XMM-SUSS4.1, XMM               & 0.19497 $\pm$ 0.00002\\

   0838-52378-0542 & 1 & 175.56190  &   5.66070 & 0.13 & 0.00 & 2MASS J11421485+0539386    & 2RXS, XMM                      & 0.10178 $\pm$ 0.00002\\

   2267-53713-0034 & 2 & 122.99395  &  15.10774 & 0.12 & 0.00 & LEDA 1475458               & 2RXS                           & 0.08355 $\pm$ 0.00002\\

   2007-53474-0294 & 3 & 161.77599  &  37.87385 & 0.10 & 0.00 & 2MASS J10470625+3752259    & 2RXS                           & 0.10193 $\pm$ 0.00003\\

   0856-52339-0010 & 3 & 210.40265  &   4.27423 & 0.10 & 0.00 & 2E 3160                    & 2RXS                           & 0.16422 $\pm$ 0.00003\\

   0531-52028-0353 & 1 & 208.68476  &   2.67756 & 0.10 & 0.00 & 1RXS J135444.8+024122      & 2RXS                           & 0.13842 $\pm$ 0.00003\\

   1233-52734-0185 & 1 & 189.45208  &   9.38978 & 0.06 & 0.00 & 2MASS J12374849+0923233    & 2RXS, XMM                      & 0.12484 $\pm$ 0.00004\\

   2610-54476-0043 & 1 & 184.62852  &  18.58286 & 0.05 & 0.00 & 2MASS J12183083+1834582    & 2RXS                           & 0.19703 $\pm$ 0.00003\\

   2507-53876-0253 & 1 & 174.73016  &  21.29640 & 0.05 & 0.00 & 2MASS J11385524+2117467    & 2RXS                           & 0.15608 $\pm$ 0.00003\\

   0306-51637-0634 & 2 & 219.26718  &   0.11807 & 0.03 & 0.00 & LBQS 1434+0020             & 2RXS                           & 0.14021 $\pm$ 0.00003\\

   0303-51615-0146 & 2 & 213.83123  &  -0.50599 & 0.03 & 0.00 & 2QZ J141519.4-003022       & 2RXS, XMM                      & 0.13451 $\pm$  0.00002\\
   0521-52326-0639 & 2 & 190.83327  &   2.88227 & 0.03 & 0.00 & 2MASX J12431998+0252562    & 2RXS, XMM                      & 0.08670 $\pm$ 0.00003\\

   0456-51910-0133 & 2 &  41.44209  &  -8.97836 & 0.02 & 0.00 & 2MASS J02454610-0858423    & 2RXS                           & 0.14823 $\pm$ 0.00002\\

   1790-53876-0564 & 1 & 192.11856  &   8.52022 & 0.01 & 0.00 & 2MASS J12482844+0831127    & 2RXS, XMM                      & 0.11858 $\pm$ 0.00005\\

   0913-52433-0213 & 1 & 206.35291  &  -2.99440 & 0.01 & 0.00 & 2dFGRS TGN201Z054          & 2RXS                           & 0.08538 $\pm$ 0.00003\\

   2510-53877-0597 & 2 & 177.73816  &  25.53028 & 0.00 & 0.01 & 2MASX J11505716+2531484    & 2RXS                           & 0.08592 $\pm$ 0.00003\\

            \hline
    \end{tabular}}
\end{table*}

\end{document}